\documentclass[a4paper,fleqn]{cas-sc}

\usepackage[numbers]{natbib}

\usepackage{graphicx}% Include figure files
\usepackage{dcolumn}% Align table columns on decimal point
\usepackage{cancel}
\usepackage{epstopdf}
\usepackage[]{graphicx}
\usepackage{bm}

\def\tsc#1{\csdef{#1}{\textsc{\lowercase{#1}}\xspace}}
\tsc{WGM}
\tsc{QE}
\tsc{EP}
\tsc{PMS}
\tsc{BEC}
\tsc{DE}
\usepackage{lineno,hyperref}

\renewcommand{\a}{\alpha}

\newcommand{\de}{\partial}

\renewcommand{\l}{\lambda}

\renewcommand{\dim}{\mathsf{dim}}

\renewcommand{\to}{\rightarrow}
\newcommand{\eps}{\varepsilon}

\renewcommand{\l}{\lambda}

\usepackage{numcompress}

\begin{document}
\shorttitle{De-localizing  brittle fracture}
\shortauthors{O. U. Salman et~al.}

\title[mode = title]{De-localizing  brittle fracture}                      
\author[1]{O. U. Salman}[orcid=0000-0003-0696-521X]
%\author[2]{G. Vitale}[]
\author[2]{L. Truskinovsky}[]
\address[1]{LSPM, CNRS - UPR 3407, Université Sorbonne Paris Nord, 93430 Villetaneuse, France}
\address[2]{PMMH, CNRS - UMR 7636 PSL-ESPCI, 75005 Paris, France}

 \begin{abstract}
Extreme localization of damage in conventional brittle materials is the source of a host of undesirable effects. We show how artificially engineered metamaterials with all brittle constituents can be designed to ensure that every breakable sub-element fails independently. The main role in the proposed design is played by high contrast composite sub-structure with zero-stiffness, furnishing nonlocal stress redistribution. The ability to de-localize  cracking in such nominally brittle systems is revealed by the fact that their continuum description is dominated by the gradient (bending) rather than classical (stretching) elasticity. By engineering a crossover from brittle to effectively ductile (quasi-brittle) behavior in  prototypical  systems of this type, we reveal the structural underpinning behind the difference between fracture and damage.  
%materials 
%differentiating between the ones, exhibiting macro-cracking-dominated fracture, and the other ones, featuring micro-cracking-dominated damage.
\end{abstract}


\maketitle

\section{Introduction}
Typically brittle materials    fail    through the development of    a macro-crack which  originates from a  microscopic flaw and advances by focusing singular  stresses near the   tip \cite{Bertram_Broberg1999-uf,noauthor_undated-vj}. 
Such   extreme stress localization  can be traced to the  nonconvexity of the  inter-atomic potential \cite{Kanninen1985-qs,Fracturett}  leading to  softening of elastic  response and attendant    loss of ellipticity of equilibrium equations \cite{Triantafyllidis:1986aa,Belytschko2003-rz}.  In  applications, the formation of a system spanning macro-crack  is often viewed as an undesirable effect  due to   low resulting toughness and  the catastrophic  character of ensuing dynamics \cite{Baker1988-uh,Swadener2002-pu}.

Many ingenious strategies for tempering brittleness and creating effective ductility have been proposed in the literature, including   toughening by microcracking \cite{Evans1981-st,Faber1988-kv},  engaging phase transformations \cite{Budiansky1983-yq,Green2018-dq} and utilizing multi-level failure mechanisms \cite{Del_Piero2001-lb,Cherkaev1995-ne}, to mention just a few. The  usual underlying  idea is to avoid the unstable crack propagation  \cite{Wiederhorn1984-ul} by creating obstacles and dissipation centers \cite{Davies1995-vv,Yang2018-ub,Borja_da_Rocha2020-vf} that can  trap  the system in metastable configurations \cite{Daniel2017-rj, Ebrahimi2019-gn}. It was shown that  such crack re-channeling mechanisms  can  be tailored to achieve high energy absorption and  that the resulting nominally brittle artificial materials can rival their  natural ductile analogs \cite{Wu2020-zj, Liang2020-dr,Radi2020-gk}.

In this paper, we propose a different approach to fracture de-localization, which can be viewed as rigidity mitigation \cite{Driscoll10813}. The underlying idea is to buttress the material's progressive softening by strengthening the \emph{nonlocal} interactions that ensure stress redistribution and prevent strain localization.  The task of transmitting such interactions is assigned to a distributed floppy metamaterial sub-structure whose elasticity is \emph{bending} dominated.

As a proof of principle, we present here the simplest prototype of such a mechanical system. Our clearly oversimplified model example aims to elucidate, by making analytically tractable, the transition from the conventional elastic behavior at small tensile loading (and no damage) to the bending-dominated elastic response at considerable tensile loading (and appreciable damage). We show that such transition ensures that instead of softening-induced strain localization, the system redistributes strain globally so that every single brittle sub-element breaks independently. The proposed design relies on the stabilization of floppy mechanical modes through bending rigidity. It has a bio-mimetic origin as it structurally imitates some known prototypes in living nature \cite{buxton2007,RevModPhys.86.995}. However, such structures with variable-connectivity can now be built artificially using the existing techniques of additive manufacturing \cite{Zadpoor2016-yr}.

Our composite design  is deliberately minimalistic as we assemble it  using the most simple  local  and nonlocal   elastic sub-structures. The local sub-structure is represented by a chain of springs with Lennard-Jones-type nonconvex potential. The nonlocal sub-structure is a zero-stiffness pantograph composed of inextensible but flexible beams connected through ideal pivots \cite{alibert:hal-00993920, Giessen:2011aa,Franciosi2019-vz}. The two sub-structures are coupled in such a way that in the initial state, where all breakable springs are intact,  the whole system is over-constrained \cite{Maxwell1864-or,Calladine1978-zc}.  As the structure is stretched, the geometrical constraints force the breakable elements to fail, and the composite mechanical system progressively transforms into an under-constrained one with dominating bending (gradient) elasticity. While we use only a particular  nonlocal sub-structure in our analysis, other  floppy  designs could be used as well, see for instance,  \cite{Schenk2014-ex} and examples in our concluding Section. A general analysis  of the  nonlocality in such systems can be found in the theory of high contrast elastic  composites \cite{pideri1997, cherednichenko2006}. Contrary to  what is known  for   usual composite materials,  higher-order derivatives  in  the  homogenized   representation of such systems,  appear already at the leading order   \cite{Camar-Eddine2003-di,boutin2013,Bacca2013-km}.% \cite{You1996-nb,Giorgio2017-xt,,Abdoul-Anziz2019-rb,DellIsola2016-tb,Alibert2015-ml,DellIsola2016-tb,Alibert2003-ny}
 
If,  in the absence of  a  floppy  reinforcement, a series  of breakable springs loaded in tension fails abruptly with a formation of a single macro-crack, e.g.,  \cite{Truskinovsky1996-xy}. Instead, we show that  the same system with a floppy   reinforcement breaks gradually and  exhibits distributed microcracking. The whole process can be interpreted as  damage  \emph{spreading}  and  viewed as a propagation of a \emph{diffuse}  front separating affine and non-affine deformation configurations. Most remarkably,  due to the presence of nonlocal reinforcement, the affine deformation is recovered at a sufficiently large   stretching with strain uniformity being \emph{rebuilt} by bending elasticity. Such re-entrant homogeneity of deformation distinguishes the proposed metamaterial structures   from the conventional  brittle solids because the latter  cannot  \emph{heal}  the acquired  non-affinity in monotone loading.
  
We show that in the continuum framework, our metamaterial structure  can be modeled as a softening elastic solid  with a strain gradient term in the energy representing bending elasticity. The ensuing continuum model takes the classical Ginzburg-Landau   form with macroscopic strain playing the role of order parameter~\cite{golubovic1989,Marconi2005-xk}. However, to describe fracture, the usual   double-well energy has to be replaced by Lennard-Jones type potential. A model of this type was considered in \cite{Triantafyllidis:1993aa}, but under a  constitutive condition  
which effectively erased the  healing  effect. Another closely related model is the  strain-gradient-regularized  damage mechanics \cite{Placidi2015-km}, though in this framework, our crucial assumption that the nonlocal stiffness is damage-independent has been so far considered  as unrealistic \cite{Le2018-jz}.  A conceptual link can also be built between our approach and models developed to describe ductile fracture in plastic solids when   an effective local energy is complemented by a  weakly nonlocal term describing strain gradient hardening \cite{Fokoua2014-gd, Fokoua2014-el}.

To elucidate the possibility of fracture delocalization \emph{patterns} 
in our  model  we also consider  a  version with  elastic background,  particularly relevant for biological applications \cite{C3SM50838B}. In this setting, the `effectively ferromagnetic' interactions, implied by bending elasticity, compete with `effectively anti-ferromagnetic' interactions brought by the elastic background \cite{Ren2000-bt}. The resulting mechanically frustrated system is shown to generate complex periodic arrangements of alternating affine and non-affine behavior.
 
If a nonlocal sub-structure is absent, the remaining local sub-structure represents a conventional brittle material that can be simulated in the continuum limit by  gradient damage model \cite{Barenblatt1993-uu,Fremond1996-ad,Bourdin2000-pc,Lorentz2003-oj,Benallal2006-yc} or any other  phase-field model of fracture \cite{Ambrosio1990-cw,Aranson2000-aj,PhysRevLett.87.045501,Eastgate2002-bo}.  
In this setting, we show that neither damage-spreading nor re-entrant behavior takes place.
Even when the elastic environment is added, the broad microcracking domains, characteristic of our metamaterial response, do not appear.  Instead, we observe in this case only the conventional pattern of highly localized macro-cracks.

Our systematic comparison of the Ginzburg-Landau type elastic model with the phase-field model shows that only the former provides an adequate description of the pantograph-reinforced breakable chain. Thus, in the absence of an elastic environment, only the former generates the characteristic \emph{isola-center} bifurcation when the nontrivial branch of equilibria separates from and then reconnects to the trivial branch at two distinct points~\cite{Golubitsky1985-qy}. Rather intriguingly, the revealed relation between these two theories   in the current context (see also \cite{Le2018-jz}) is rather similar to the one between the Foppl-von Karman and the fully nonlinear 3D theories of wrinkling in stretched elastic sheets~\cite{Li2016-tq}. The fundamental fact that the phase-field model misses the crucial re-entry effect shows that the penalization  of the inhomogeneity through gradients of damage is a much weaker form of regularization than through the gradients of strain because only the latter  can ensure  the affine response even in the case of complete degradation of the conventional elasticity.

The rest of the paper is organized as follows. In Section 2, we introduce the discrete model of the pantograph-reinforced breakable chain and show that fracture in this system is de-localized. In Section 3, we build a continuum version of the same system and show that its mechanical response reproduces its discrete prototype's mechanical behavior. In Section 4, we study the case when the continuum model is constrained by an elastic environment showing the emergence of regular patterns with alternating affine and non-affine strain configurations. In Section 5, we compare (at continuum level) the behavior of pantograph-reinforced and non-reinforced brittle systems. In the final Section 6, we  present elementary  examples  of  2D systems with fracture  de-localizing floppy substructures  and summarize our results.

\section{The design idea} 
\begin{figure}[hbt!]
 \centering
\includegraphics[scale=0.2]{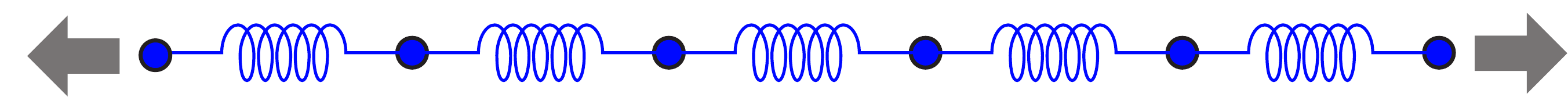}
\caption{The simplest (not reinforced) mass-spring  chain.  Particles are connected in series by breakable springs and the whole system is loaded in tension. }
\label{fig1a}
\end{figure}

To motivate further developments, consider a conventional mass-spring chain constrained to remain on a straight line, see Fig. \ref{fig1a}. The goal of this basic pre-model is to mimic the mechanical behavior of a softening nonlinear elastic material.  To this end, we assume that the springs are  `breakable' and that their mechanical response is described by a non-convex elastic potential of Lennard-Jones type. 

\begin{figure}[hbt!]
 \centering
\includegraphics[scale=0.4]{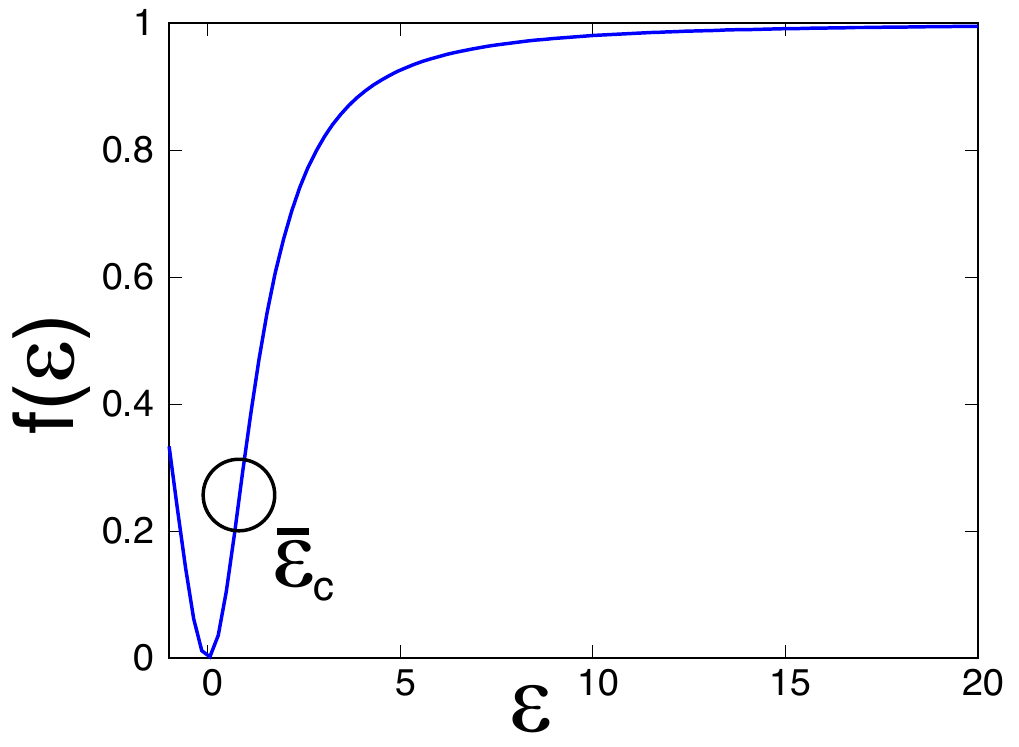}
\caption{The elastic potential of  Lennard-Jones-type describing the mechanical response  of a `breakable' spring loaded in tension.  The compression range is not  shown because it is not used in this study.  The inflection point  $\bar {\eps}_c$ marks the instability of  the affine response for the simplest   chain with nearest neighbor interactions  shown in Fig. \ref{fig1a}. }
\label{fig1f}
\end{figure}

Define the horizontal displacement $u_{i}=x_i-x^0_i$ of the   mass point with index $i$ where $x_i$ is its actual coordinate and $x^0_i=ai$ is its reference coordinate. The  energy of this system can be  written in dimensionless  form  
\begin{equation} 
E_S({\bf u}) = a\sum_{i=0}^{N-1} f\left( \frac{u_{i+1}-u_{i}}{a}\right).
\label{meq0}
\end{equation}
where  $a=1/(N-1)$ is  dimensionless reference  length.   For  our numerical illustrations,  where we  deal  exclusively with tension, it will be sufficient to   use  an  analytically convenient expression  for the elastic potential  $f(\eps_i) = \eps_i^2/(2+\eps_i^2)$ where  we introduce the discrete strain $\eps_i= (u_{i+1}-u_{i})/a $,  see Fig. \ref{fig1f}. 
 
Suppose  next that the chain  is stretched  quasi-statically in a hard device so that $u_0=-\bar \eps/2$ and $u_{N-1}= \bar \eps/2$, were $\bar \eps > 0$ is the average strain which   plays the role of  loading parameter. To find the macroscopic response we  need to solve for  each value of   $\bar \eps$ the equilibrium  equations 
 \begin{equation}  \partial{E_S }/ \partial u_i=0\text{ , } 1<i<N-2.\label{eq:discrete}
 \end{equation} 
% with $1<i<N-2$ 

Knowing that   continuous  branches of equilibria can terminate, we need to prescribe the  branch switching strategy defined by the dynamic extension of the model. Such extension  should  ensure that the system  re-stabilizes after an instability  in a dissipative way and in quasi-static setting reduces to the   selection of a new locally stable equilibrium branch with necessarily lower energy. In this paper,  we  will compare   two  dynamic strategies.

Having the  \emph{structural mechanics} applications in view, we should be choosing the new equilibrium branch using the local energy minimizing (LEM) criterion which mimics  the  zero viscosity limit of an overdamped viscous dynamics. Under  this protocol,  the quasi-static loading   will maintain    the system in a metastable state (local minimum of the energy \eqref{meq0}) till it ceases to exist and then, during an isolated switching event, select the  new equilibrium branch using the steepest descent algorithm~\cite{Puglisi2005-lg}. 

With  the \emph{ biomechanical } applications in view, where   temperatures are different from zero and the energy scale of thermal fluctuations is comparable to the existing energy barriers, we  also discuss the global energy minimizing (GEM)  strategy.  It implies that at each value of the loading parameter, the system is able to  minimize the energy globally.   Physically, this branch selection strategy indicates  the parametric  thermal equilibration with the subsequent  zero-temperature limit.  While for macroscopic engineering structures, the global energy minimization does not make much sense in the context of fracture, at  sufficiently small scales (encountered, for instance, in cells and tissues)  and over sufficiently large times,  thermal fluctuations can be thought as exploring  enough of phase space to make global minima relevant.

\begin{figure}[hbt!]
 \centering
\includegraphics[scale=0.28 ]{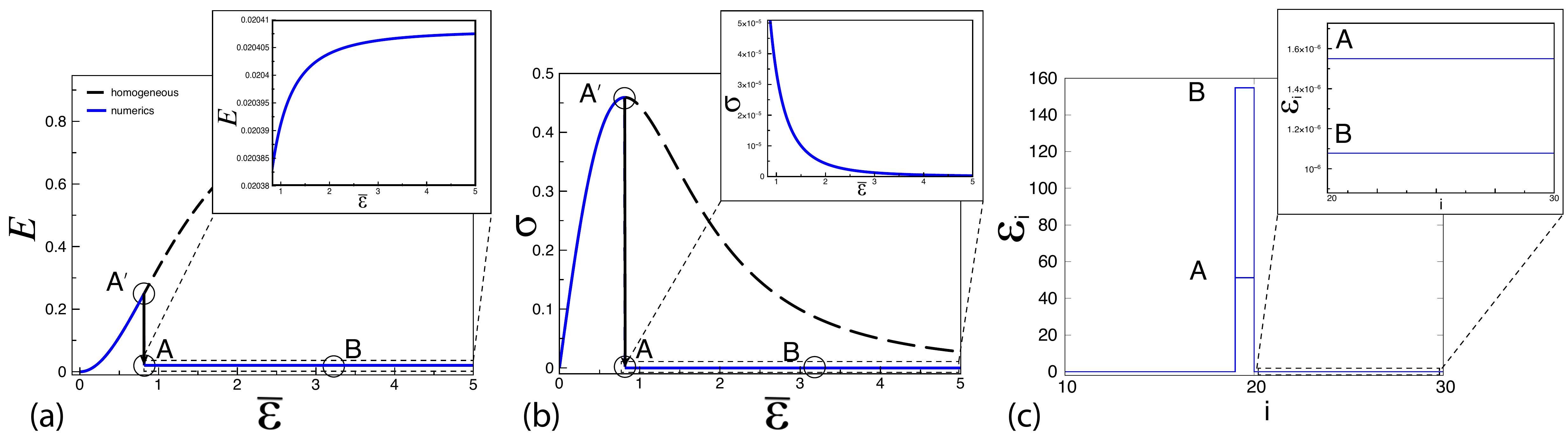}
\caption{The LEM   response of a   mass-spring chain with breakable springs subjected to quasi-static tensile loading in a hard device:  (a) the equilibrium elastic energy $\bar E_S(\bar\eps)$ and, (b) the equilibrium stress  $\sigma(\bar\eps) =d \bar E_s/d\bar\eps$, (c) the equilibrium strain profiles $\eps_i(\bar\eps)$, $i=1,...,N$. Hypothetical affine response is compared with numerical simulations following the LEM protocol. Insets: (a,b) the response of the broken chain beyond point  A, (c) the state of the unloaded springs in points A and B. Here $N=50$.} 
\label{fig2m}
\end{figure}
 
While for the system with energy \eqref{meq0} both LEM and GEM responses can be studied \emph{analytically},  we  use here this simple case to detail our numerical approach. For instance, in Fig. \ref{fig2m} we illustrate the classical brittle fracture with ultimate (lattice scale) strain localization. Here we  apply the LEM strategy while stretching the chain with $N=50$. The incremental energy minimization is based on  L-BFGS algorithm~\cite{Bochkanov2013-lk} (imitating gradient flow), which builds a positive definite Hessian approximation for \eqref{eq:discrete} allowing one to make a quasi-Newton step lowering the total energy~\cite{Chan1982-xj}. Such iterations continue till the increment in the energy becomes sufficiently small.  We then use the obtained approximate solution as an initial guess $\bf w$ to solve, using LU factorization~\cite{Sanderson2016-ht}, the linear equations for the correction $ \Delta\bf w$
%around such an initial guess for the displacement field ${\bf w}^a +  \text{d}{\bf w}^a$ yields a system of linear equations:
\begin{equation}
\label{newton}
 K_{ij} \Delta w_j+\Delta f_i =0,
\end{equation}
where  
 $K_{ij}=\partial^2 E_S/ \partial u_i\partial u_j$
is the discrete stiffness matrix, and $\Delta  f_i 
%= \partial E_S/ \partial u_i
$ are the bulk forces.
The displacement field is updated in this way till   the gradient norm  of  Eq.~\ref{eq:discrete}  is  smaller than  $ 10^{-8}$  which  furnishes the actual solution of the problem. The loading is performed  by  monotonically increasing the value of  the  displacements  of boundary nodes  $u_0=-\bar\eps/2$ and $u_{N-1}=\bar\eps/2$
 in increments of $10^{-6}$. The Hessian matrix $ K_{ij} $ is also used to assess the stability  of the obtained equilibrium configurations. To determine the GEM path we simply chose at each value of $\bar\eps$  the equilibrium configuration with the lowest energy.

In Fig. \ref{fig2m} we show separately  the equilibrium macroscopic energy, $\bar E_S(\bar\eps)$,   the  equilibrium macroscopic stress $\sigma(\bar\eps) =d \bar E_S/d\bar\eps$ and the equilibrium distribution of the microscopic strains  $\eps_i$ in individual springs. The homogeneous (affine)  configuration  remains locally stable till the   value $\bar\eps=\bar \eps_c$ is reached, where $\bar \eps_c$ is defined by the instability condition  $ \partial^2 f/ \partial \eps^2 =0$, see Fig. \ref{fig1f}. As the homogeneous  state  becomes unstable, the stress drops  to almost zero  and then continues to  diminish further as the loading parameter increases, see the inset in Fig. \ref{fig2m}(b).  During the stress drop,  the strain abruptly localizes at the scale of the lattice, see Fig. \ref{fig2m}(c).  The location of the  ensuing macro-crack is accidental  and was chosen by  an  initial imperfection. Subsequent loading increases the crack  opening while further unloading the rest of the sample, see Fig. \ref{fig2m}(c).  Under  the GEM protocol, similar isolated macroscopic crack  forms  before the point $\bar \eps_c$ is reached, see for instance, \cite{Truskinovsky1996-xy}, however, the  subsequent growth of this crack  follows the same equilibrium branch as in LEM case. The main difference is that  under the LEM strategy, the dissipation during the  brittle failure process  is   finite,  while under the GEM strategy the dissipation is identically   zero.

After the major stress drop the subsequent loading does not create additional damage and the response reduces  to the  increase  of the amplitude  of the localized strain. Note also that during the stress drop, the system does not fully unload because some  (weakening) elastic interaction between the newly formed crack lips always exists.
 A 1D model of brittle fracture where stress drops to zero while the boundary layers near the lips remain have been recently proposed in~\cite{Rosakis2021-qk}.

\begin{figure}[h!]
 \centering
\includegraphics[scale=0.17 ]{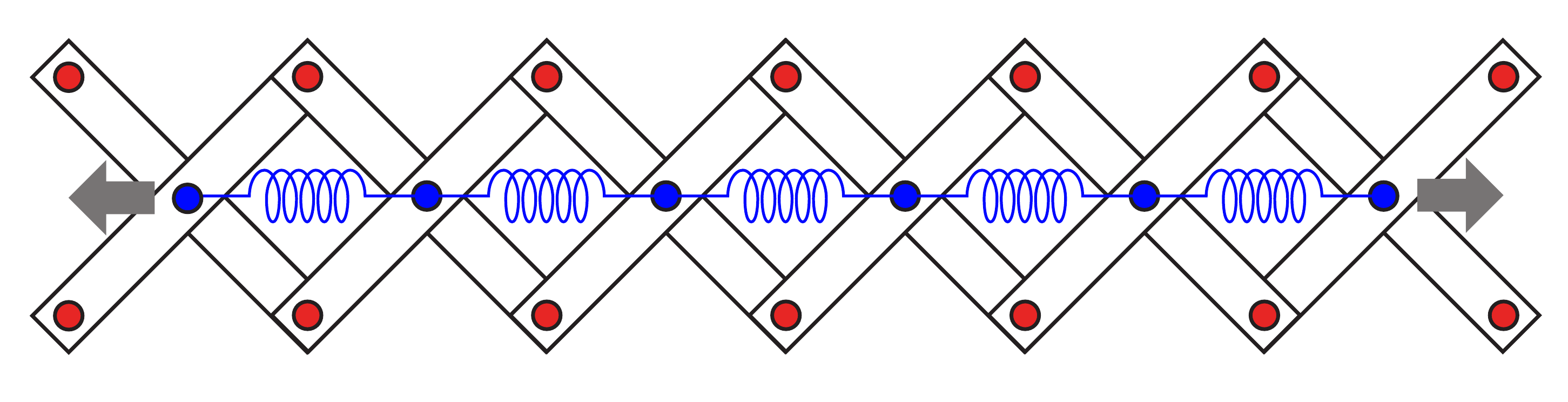}
\caption{The reinforced  chain of breakable springs. The  pantograph floppy frame constitutes a sub-structure. The system is loaded in tension.}
\label{fig1b}
\end{figure}

To de-localize brittle fracture, we now reinforce the series connection of breakable springs by adding a \emph{sub-structure} whose role  is to ensure that the strain is  uniformly redistributed. The simplest sub-structure of this type is a \emph{pantograph} frame made of inextensible but bendable beams connected by ideal pivots, see Fig. \ref{fig1b}. While it has zero longitudinal macroscopic stiffness, (it does not resist affine deformations)  the non-affine longitudinal deformations remain energetically penalized due to the finite bending rigidity of individual beams.  If we again assume   that  the composite system shown in  Fig. \ref{fig1b}    is    constrained to remain on a straight line,  we can    write   the elastic  energy of the bending beams in the form 
\begin{equation} 
E_B({\bf u}) =  a\sum_{i=1}^{N-1} \frac{\lambda_1^2}{2}\left( \frac{u_{i+1}+u_{i-1}-2u_{i}}{a^2}\right)^2,
\label{meq00}
\end{equation}
where $\lambda_1 $ is a dimensionless length proportional to $a$ with the  coefficient depending on the  bending stiffness of the  beams  \cite{Alibert2003-ny}.
%  of length $l=\sqrt{2}n^{-1}$ and it is related to the standard flexural stiffness $K$ of an inxtensible beam of length $l$ through $\lambda_1^2=\frac{3\sqrt{2}K}{2}$. 
The total energy of the  composite system  is then 
\begin{equation}
 E({\bf u}) =E_S({\bf u})  + E_B({\bf u}),
\label{meq000}
\end{equation} 
where $E_S({\bf u})$ is given by \eqref{meq0}. We will again load  the system   in a hard device with $\bar \eps$ serving as  the loading parameter. No other  constraints are imposed, making the ends of the reinforced structure effectively `moment free' ~\cite{Charlotte2002-fr}.
 
\begin{figure}[hbt!]
 \centering
\includegraphics[scale=0.3 ]{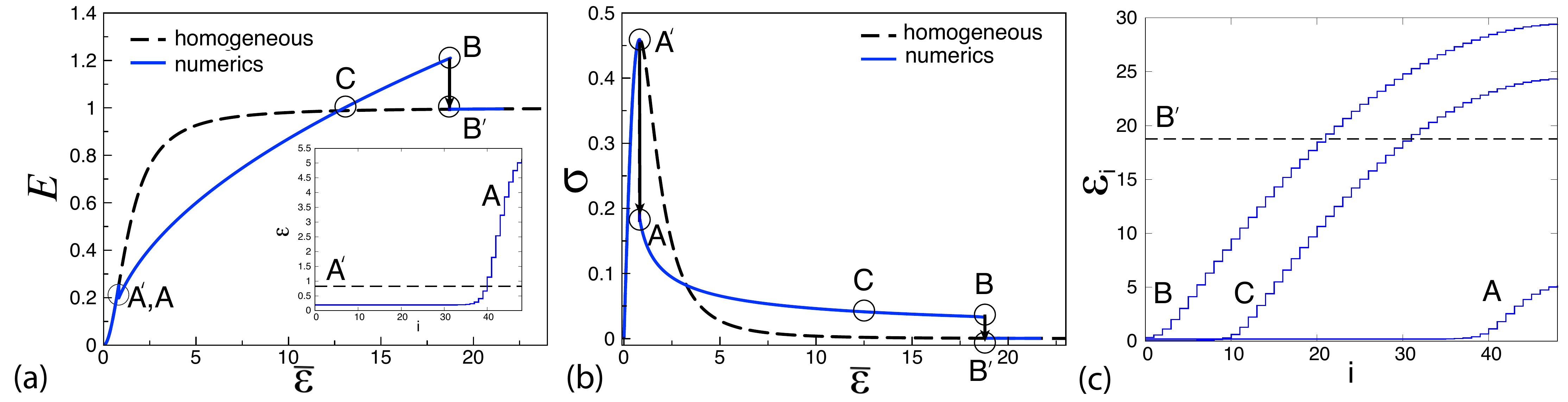}
\caption{Mechanical response  of the pantograph reinforced chain subjected to tensile loading in a hard device and following LEM dynamic protocol: (a) the equilibrium elastic energy $E(\bar\eps)$; (b) the equilibrium stress  $\sigma(\bar\eps)$; (c) typical strain profiles. Inset details  the transformation  the homogeneous (affine) state  A'  into an inhomogeneous (non-affine) state A which takes the form  of the nucleation of distributed damage at the boundary of the sample. The affine state is recovered when the inhomogeneous state  B, describing a developed damage zone, transforms back into the homogeneous state B'. This re-homogenization transformation would have happened at C if the system followed the global minimum path. Here, $N=50$, $\lambda_1=0.02236$.} 
\label{fig:bending}
\end{figure}

The mechanical response of the pantograph-reinforced chain under the LEM protocol is illustrated in Fig. \ref{fig:bending}. To obtain the response curves, we used the same numerical approach as in the case of the non-reinforced chain. The energy minimizer is again affine at sufficiently small values of the loading parameter.  At a critical value of the load (point A') the homogeneous state becomes unstable, meaning that the corresponding local minimizer ceases to exist. The local energy minimization, imitating overdamped dynamics, produces  the drop in stress  and  brings the system to a new local minimum (point A). However, instead of  an \emph{isolated} localized crack, the failure process now produces a \emph{diffuse} nucleus of strain non-affinity. The strain profiles, see Fig. \ref{fig:bending}(c), suggest that  the transition A'$\to$A leads to  the formation of a damaged (micro-cracked) zone.  The fact that it forms on one of the  boundaries of the sample is a consequence of the zero moment boundary condition which effectively weakens the boundaries.  

As the loading continues,  the damage spreads through the structure, see the configurations C and B in Fig. \ref{fig:bending}(c). The increase of the total energy accompanies the process of successive microcracking while the stress progressively declines, see Fig. \ref{fig:bending}(b). The  `de-localized' failure zone advances into the sample from A to B as more and more springs get effectively broken, see Fig. \ref{fig:bending}(c). Note that the advancing zone of strain non-affinity does not have a sharp front but, instead, propagates as a diffuse wave.

According to LEM dynamics, another critical value of the loading parameter is at point B where  the damage instantaneously spreads through the remaining, still-intact part of the sample, and the deformation becomes affine again all over the sample (transition B $\to $B'). This is the second dissipative event as the  energy,  in addition to  stress, drops abruptly. The newly acquired homogeneous response remains energy minimizing at larger strains, see Fig. \ref{fig:bending}(a-c). Note that the total dissipation in this composite system, measured by the area under the  resulting stress-strain curve shown in Fig. \ref{fig:bending}(b), is higher than in the case of the simple non-reinforced chain, see Fig. \ref{fig2m}(b).

The GEM response of the pantograph-reinforced  system will be roughly similar with the transition A'$\to$A taking place at smaller values of the loading parameter chosen by the corresponding Maxwell condition. Similarly, the re-stabilization of the affine state will take place earlier than under  LEM  protocol, at the point C, rather than B,  see Fig. \ref{fig:bending}(a-c). Despite these two abrupt stress drops along the GEM path, no energy is dissipated in  such a process with all the work of the loading device  absorbed by the system.

We have seen that if the simplest  LEM model of a breakable chain   produces   conventional    localized  fracture with  an abrupt drop of stress and low dissipation, the LEM model of  a pantograph-reinforced chain shows  an unusual   de-localized  fracture with  a gradual decrease of stress and  higher energy dissipation.  Instead of breaking into  two pieces,   the pantograph-reinforced chain    \emph{fragments} uniformly   into $N$ equal pieces.  The  brittle response of the original system is then replaced by a  ductile one with  stable  \emph{softening} behavior  and  larger effective toughness. We remark that the advantages  of embedding brittle elements into a compliant matrix are  well known in both engineering (fiberglass) and biology (extra-cellular matrix). A certain peculiarity of our toughening mechanism is that it produces overall softening behavior instead of  the more conventional hardening utilized,  for instance,   in ceramic matrix composites  \cite{Curtin1998-fs}.

\section{Continuum analog} 

The model shown in  Fig. \ref{fig1b} was designed as a prototype of an inherently discrete system (metamaterial). However, such nonlinear discrete systems are not analytically transparent  and therefore the origin of  their  unusual mechanical behavior is not apparent. Some theoretical insights can be obtained if we consider their mathematically  more  tractable continuum analogs.

Suppose that the (dimensionless) lattice parameter $a$ is sufficiently small. We can then look for  a continuum model which, in the limit $a \to 0$, is  asymptotically equivalent to our discrete model. Since the formal, scale-free asymptotic limit of the discrete theory produces a degenerate model \cite{Truskinovsky2010-st,Braides1999-mp}, we should aim at \emph{(quasi) continuum} description preserving the lowest order terms in the small parameter $a$.

Approximations of  different order, based on the idea of  $\Gamma-$ convergence, were discussed in \cite{Braides2008-vt}. However, since here we are interested in local minimization of the energy, $\Gamma-$ limits are not adequate, and we need to use instead the parallel approach based on the computation of point-wise limits \cite{Blanc2002-ss,Charlotte2008-qx}. The simplest low-order quasi-continuum  approximation of this type can be obtained if  we use  formal asymptotic expansions of the linear finite difference  operators in  $a$.
% must necessarily survive because  the  formal asymptotic  limit gives a  degenerate theory \cite{Truskinovsky2010-st}. Therefore, we need to  construct a quasi-continuum with a  finite length scale $a$. 
Using these ideas we  formally replace the discrete energy functional $ E({\bf u}) $ in \eqref{meq000} by its (quasi) continuum analog 
\begin{equation}
\label{eq1}
E(u) = \int_{0} ^1\left( f(\eps)+\frac{\lambda_1^2}{2} \eps'^2 \right)dx,
\end{equation}
where $\eps(x) =u'(x) $ is the continuum strain variable and $u(x)$ is the corresponding displacement field. The energy density in \eqref{eq1} maintains the additive structure of its discrete analog with the first term representing the stretching energy of the breakable springs and the second term describing the bending energy of the pantograph substructure. The 'redressed' parameter $\lambda_1$, whose exact value will not be of interest in our qualitative study, is  assumed to be strain independent given that the beams can be viewed as much stiffer than springs. It  brings into the ensuing (quasi)continuum model a dimensionless length scale which does not fade away with loading as in models of un-supported mass-springs chains \cite{Triantafyllidis:1993aa}. For numerical illustrations we continue to use the particular function $f(\eps) = \eps^2/(2+\eps^2)$. 

To model  the system loaded in a hard device, we again set 
$
	 u(0) = - \bar{\eps}/2 $, $u(1) = \bar{\eps}/2
$ 
where
$\bar{\eps}>0$ 
is the imposed strain. Given that the boundaries  are  moment free, we also use  the natural higher order boundary conditions 
$u''(0) = u''(1) = 0.$  
% Note that  while in the stretching dominated phase, where the stiffness $\de^2 f>0$,  the correlation length $\sim \lambda/|\de^2 f_0|$ (\textbf{check}) is finite,   in the bending dominated phase,  where $\de^2 f \rightarrow 0$, it can grow to infinity \cite{Charlotte2008} . 
%(\textbf{min of this energy should be at zero otherwise the meaning of $u$ has to be changed}) 
Under these assumptions   the homogeneous (affine) configuration 
%\begin{equation}\label{eq1_1}
$u^0(x) = (\bar{\eps}/2)(2x - 1)$
%\end{equation}
  is   an equilibrium state at all values of the loading parameter $\bar{\eps}$. 
  
\begin{figure}[hbt!]
\begin{center}
\includegraphics[scale=0.65]{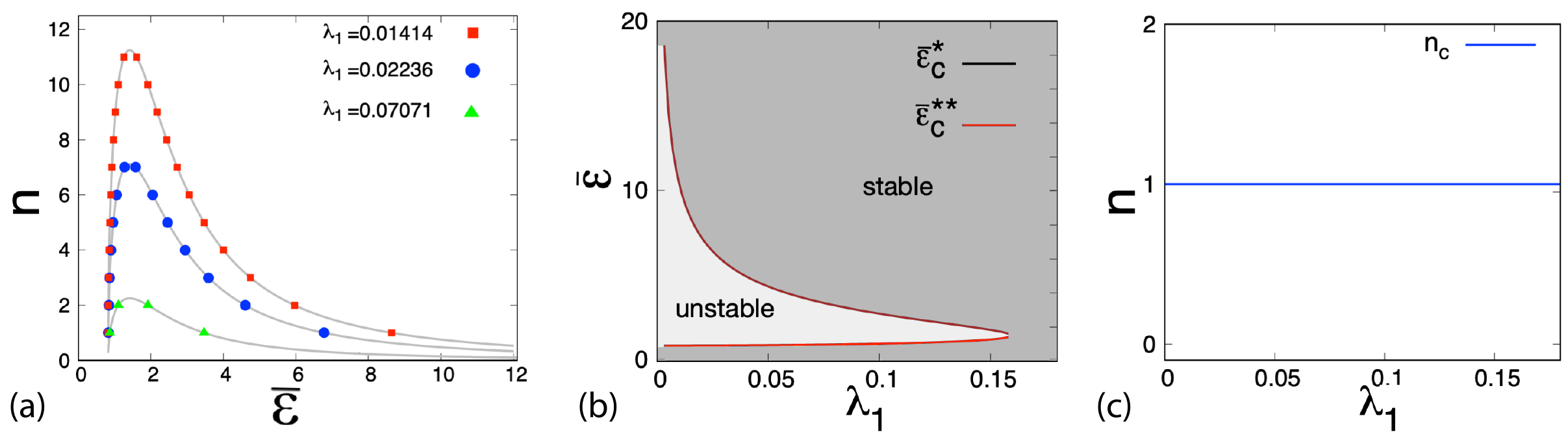}
\caption{Linear stability boundaries for  the homogeneous state in the continuum model of pantograph-reinforced chain: (a) bifurcation points, (b) critical strains  $\bar{\eps}_c^*$ and $\bar{\eps}_c^{**}$ bounding the unstable domain (c) critical wavenumber  $n_c$ vs $\lambda_1$.  }
\label{fig2_2}
\end{center}
\end{figure}

 Due to the softening  nature of the energy density $f$,  the homogeneous configuration  can be  expected to become unstable in tension. To find the critical value of the loading parameter   we need to study  the  linear problem for the   perturbation $s(x)=u(x)- u^0(x)$   
\begin{equation}
\label{eq2}
 -\lambda_1^2 s'''' +  \frac{\de^2 f}{\de\eps^2} (\bar{\eps}) s'' = 0,
\end{equation}
with the boundary conditions  
	$s(0) = s(1) = s''(0) = s''(1) = 0.$ 
The system becomes linearly unstable  when the second variation of the energy \eqref{eq1}  loses its positive definiteness and the problem \eqref{eq2} acquires a nontrivial solution.  The largest eigenvalue of the linear operator with constant coefficients in the left hand side of \eqref{eq2} is $ -  \lambda_1^2 \pi^4-(\de^2 f/\de\eps^2) (\bar{\eps})\pi^2$ and the corresponding eigenvector   is $s(x)\sim\sin( \pi x)$ \cite{Lifshits1985-vc,Lifshitz1985-hm}. Therefore, the  homogeneous solution is stable for  $(\de^2 f/\de\eps^2) (\bar{\eps})  > - \lambda_1^2 (\pi)^2$ and   the loss of stability  takes place at the smallest $\bar{\eps}$ such that $(\de^2 f/\de\eps^2) (\bar{\eps})  = - \lambda_1^2 \pi^2$. The higher order modes $s(x)\sim\sin(n \pi x)$ with $n>1$ bifurcate at the values of the loading parameter satisfying 
 \begin{equation}
 \label{eq3}
   \frac{\de^2 f}{\de\eps^2} (\bar{\eps})  = - \lambda_1^2 (n\pi)^2.  
 \end{equation}
%with unstable mode  $s(x)\sim\sin(n \pi x)$.  
 The corresponding \emph{stability boundaries}, representing  solutions of  \eqref{eq3} at different values of $n$ and $\lambda_1$, are illustrated in Fig. \ref{fig2_2}(a,b) for   the  case of our  special $f(\eps)$.  In Fig. \ref{fig2_2}(a,b) we show that, independently of the value of the parameter $\lambda_1$, the wavelength of the critical perturbation   always corresponds to $n_c=1$.
 
Note a remarkable feature of the stability diagrams shown in  Fig. \ref{fig2_2}(a,b). If $\lambda_1$ is sufficiently small, the homogeneous (affine) configuration is  stable in the two disconnected domains:  when the applied stretch is sufficiently small  $\bar{\eps} \leq\bar{\eps}_c^*$ and when it is  sufficiently large $\bar{\eps} \geq\bar{\eps}_c^{**}$, with the same critical mode number   ($n_c = 1$)  responsible for  both instabilities, see Fig. \ref{fig2_2}(c).  This observation     points  to the existence  of the  re-entrant behavior  characteristic for the isola-center   bifurcations \cite{dellwo1982,Golubitsky1985-qy,
Healey2013-ou}.  When the dimensionless parameter $\lambda_1$  is  sufficiently  large (large bending modulus or small system size), the affine configurations are stable  in  the whole range of  loadings.   In such `overconstrained' regimes,  failure becomes `dissipationless,'  taking place gradually and uniformly  throughout  the whole system.

The metastable   non-affine  configurations can be found by solving the   nonlinear equilibrium equation
 \begin{equation}
\label{eq4}
-\lambda_1^2  u'''' +    \frac{\de^2 f}{\de \eps^2} (u')  u''  = 0,
\end{equation} 
with the chosen boundary conditions for $u(x)$. The whole set of solutions  of  the nonlinear equation \eqref{eq4} can be obtained  in quadratures, e.g.,~\cite{Lifshits1985-vc,Lifshitz1985-hm,Truskinovsky1995-ns}. For the general problem with  an arbitrary convex-concave potential $f(\eps)$, different equilibrium branches can be parametrized by the unstable wave number $n$ at the point of bifurcation from the trivial homogeneous solution;  the latter itself can be associated with $n=0$.   One can show that all    branches   $u_n(x)$   with $n >1$ are all unstable~\cite{Novick-Cohen1984-qm,Lifshits1985-vc,Lifshitz1985-hm}. It can be also shown that the branch with $n=1$ which bifurcates from the trivial homogeneous branch of equilibria at  $\bar{\eps}_c^*$ does it subcritically  and that it  reconnects to it  at $\bar{\eps}_c^{**}$ also subcritically \cite{Lifshitz1985-hm}.

\begin{figure}[h!]
\centering
  \includegraphics[scale=0.38]{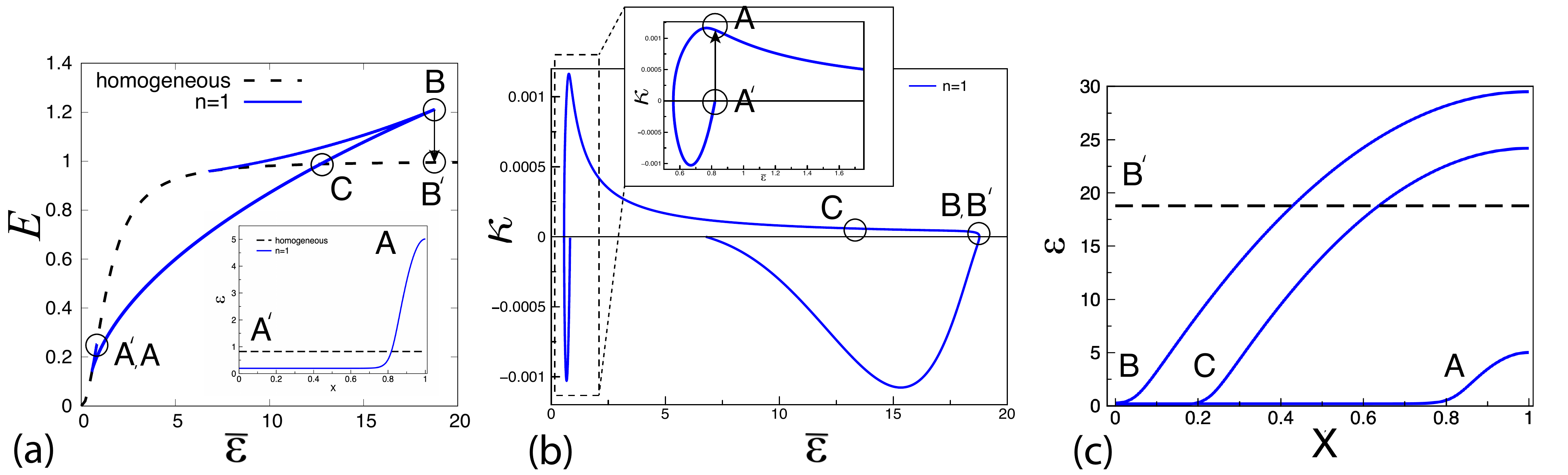}
  \caption{The LEM mechanical response in tensile loading for the continuum model of a pantograph  reinforced   chain: (a)  equilibrium energy-strain relation  for the  branch $n=1$ with the lowest energy, (b) smallest eigenvalue of the second variation for the mode $n=1$  as a function of the loading parameter $\bar\eps$.  The inset in (a) shows the nucleation of the diffuse damage zone near the boundary of the sample (transition A'$\to$A).    In point B  the non-affine state    transforms into the affine  state (transition B$\to$B').  Along the global energy minimization path (GEM response) the   re-entry type transformation would   take place in point C. Here $\lambda_1=0.02236.$}
\label{energy-1a}
\end{figure}

\begin{figure}[h!]
\centering
  \includegraphics[scale=0.35]{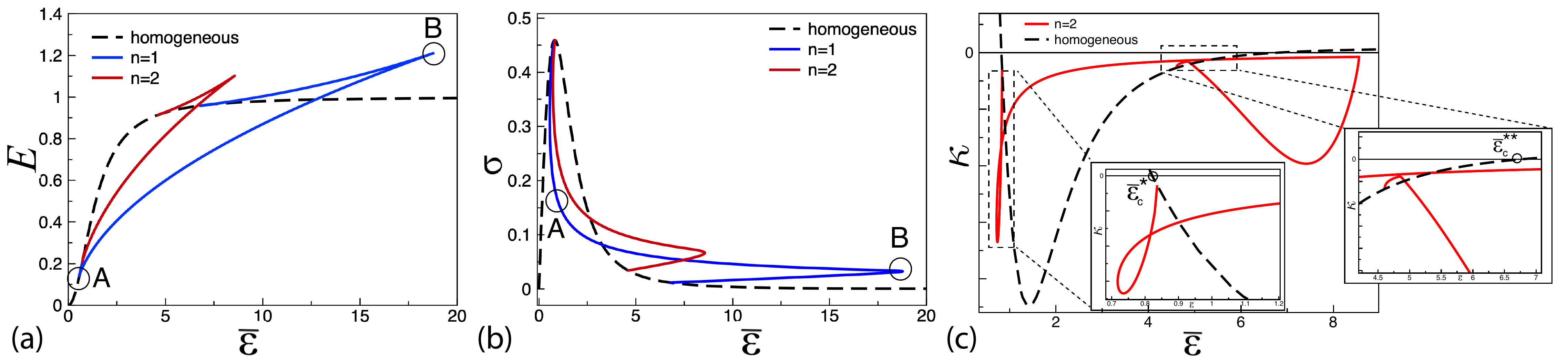}
  \caption{The mechanical response in the continuum model of a pantograph  reinforced  chain : (a)  equilibrium energy-strain relations showing  the two lowest energy branches, (b) the corresponding stress-strain  relations, (c) minimal eigenvalues of the second variation for branches with $n=0$ (homogeneous) and $n=2$ (unstable). }
\label{energy-1b}
\end{figure}

These general observations are confirmed numerically in the case of   our special  potential $f(\eps)$, see Fig. \ref{energy-1a}(a) and Fig. \ref{energy-1b}(a,b).  Equilibrium branches were found   using a pseudo-arclength continuation technique, implemented in the software AUTO  \cite{doedel08auto-07p}. It solves the nonlinear equation Eqs. \ref{eq4} with the relative end displacement treated as a continuation parameter. 
%The program follows the main branch of solutions, determines the bifurcation points, then calculates the bifurcating branches. 
To discretize the boundary-value problem, it uses collocation with Lagrange polynomials, and in our simulations, we had $N=200$ mesh points with $N_c=5$ collocation nodes and activated mesh adaptation. 
%First, we checked that  AUTO finds the bifurcation points that we calculated analytically through the linear stability analysis, (see Figs. \ref{fig2_2} and \ref{fig2_3}). Second, we followed the bifurcated branch until it reconnects to the homogeneous branch. 
%textbf{All branches with $n \neq 1$ are unstable. ( check numerically for $n=2$)}
To study stability we checked numerically  the   positive definiteness of the second-variation 
\begin{equation}
\delta^2 E(u)(v,v)=\int_0^1[(\de^2 f /\de\eps^2) (u')v'v'+\lambda_1^2v''v'']dx,
\label{hessian11}
\end{equation}
where $v$ are the  test functions respecting the boundary conditions. We  discretized  the integral \eqref{hessian11} to construct the stiffness matrix ${\bf K}$  and then investigated numerically the sign of the minimal eigenvalue $\kappa$ of the corresponding  finite quadratic form~\cite{Sanderson2016-ht}. To this end, we used one-dimensional finite elements based on third-order polynomial shape functions containing four unknown constants (cubic Hermite interpolation)~\cite{Liu2013-sj}. This implies that four shape functions were used in each two-node element (4 degrees of freedom), and we used a uniform mesh with an element size $h_e=1/1000$. The discrete solution $u'(x_i)$  provided at discrete nodes $x_i$ by AUTO was first interpolated using B-spline basis function of degree 3 ~\cite{SPLINTER} and then used to calculate the integral \ref{hessian11} using a three-point Gauss integration scheme; the fixed boundary conditions were imposed by removing from the stiffness matrix ${\bf K}$ the row and columns at $x=0$ and $x=1$. As a result, the second variation \eqref{hessian11} was approximated by a finite sum with 
$
K_{ij}=\int_0^1[(\de^2 f /\de\eps^2) (u'){\mathcal N}'_i {\mathcal N}'_j+\lambda_1^2 {\mathcal N}''_i {\mathcal N}''_j]dx,$
%\label{hessian11_stifsness}
%\end{equation}
where ${\mathcal N}_i(x)$ is the shape function of  node $i$~\cite{Liu2013-sj}.

In   Fig.~\ref{energy-1a}(a) we show the macroscopic energy-strain relation for the  system following the LEM protocol.  The non-affine branch with $n=1$ bifurcates from the trivial branch     at point A'. Since the bifurcation is subcritical, the system jumps   from the homogeneous state A' to the inhomogeneous state A located on the only other stable equilibrium branch (with $n=1$), see the inset in  Fig.~\ref{energy-1a}(a). 
During this abrupt transition, the stress drops, and the energy is dissipated, see Fig. \ref{energy-1b}(b). As the loading parameter increases, the system follows the $n=1$  branch till it reaches the turning point B.  From there the system abruptly returns to the homogeneous (affine)  branch  $n=0$ as a result of a dissipative transition  accompanied  with another  stress drop, see Fig. \ref{energy-1b}(b). Further loading preserves the affine nature of the strain configuration which has recovered its stability.  In  Fig.~\ref{energy-1a}(b) we show the strain dependence of the lowest eigenvalue  of the second variation $\kappa$ for the equilibrium branch with $n=1$. As we see, this branch   bifurcates from the branch with $n=0$ at the  point A' as an unstable one. However, the non-affine state A,   where the  LEM solution  jumps,   is  stable as the corresponding $\kappa>0$. The branch $n=1$ loses stability  again at the turning point B where $\kappa=0$.  

We remark that the \emph{global} structure of the nontrivial branch $n=1$  is similar to the one obtained in the Ginzburg-Landau model with a double-well potential, e.g.  \cite{Lifshits1985-vc}. The same topological structure of the bifurcation diagram is observed  in a model with  a single-well potential because  the separation and the reconnection of the nontrivial branch $n=1$ with the trivial branch $n=0$ takes place outside the energy wells, in the spinodal region,  which is effectively present in both types of theories.

The structure of the associated local energy minimizers  is illustrated  in  Fig.~\ref{energy-1a}(c). The abrupt nucleation of the  first domain of non-affinity   takes place at the point A' corresponding to $\bar{\eps}_c^*>\bar{\eps}_c$.  In  the  linear regime, the   unstable bifurcating mode  has a characteristic  size of the system,  but in  the  nonlinear regime,  it  partially localizes near one of the   boundaries  (transition A'$\to$A).   As the applied strain  $\bar{\eps}$ increases,  the  non-affine   state of  distributed damage  proliferates towards the other boundary  of the sample;  note that it  remains   diffuse   as  successive springs continue to  break. Observe \emph{broad}  transition layers, separating the non-affine  domains  which contain    broken springs  (domains where   $ \de^2 f/\de\eps^2 <0$)  and  the dominating elasticity is of the bending (gradient) type,  from the affine  domains    which contain  intact  springs  (domains where  $\de^2 f/\de\eps^2 >0$)  and  the dominating elasticity is   of the classical  'stretching'  type.  The fully affine configuration is recovered through the  discontinuous event which marks  the complete annihilation of the stretching  dominated  domain (transition B$\to$B'). 

Since branches with $n=0$ and $n=1$ are the only  ones containing stable fragments, the GEM path can also be read off our Fig.~\ref{energy-1a}(a). Thus, the global energy minimizing transition from $n=0$ to $n=1$ branch takes place slightly before point A' while the reverse transition occurs at point C.  It involves the instantaneous breaking of almost half of the springs and precedes considerably the analogous transition in the LEM model taking place at the point B, see Fig.~\ref{energy-1a}(c).

An  unstable equilibrium branch with $n=2$, which does not participate in either LEM, or GEM paths,  is illustrated in Fig. \ref{energy-1b}.  It describes saddle points but exhibits similar  isola-center bifurcation.   The instability of this branch  is illustrated  Fig. \ref{energy-1b}(c) where we show  the strain dependence of the lowest eigenvalue of the second variation $\kappa$ for branches with $n=0$ and $n=2$. Note that since the $n=2$ branch bifurcates from the trivial state as the second eigenvalue of the second variation becomes negative, the curve $\kappa(\bar \eps)$ in Fig. \ref{energy-1b}(c) corresponding to $n=2$ originates   on $n=1$ branch rather than  $n=0$ branch. It is also clear from Fig. \ref{energy-1b}(a) that the branch $n=2$ corresponds to a higher energy level than the branch $n=1$  and, therefore, independently of the local stability analysis, it cannot make appearance  in principle along  the GEM path.

Our analysis shows that the behavior of the (quasi)continuum model is qualitatively  similar to the behavior of the corresponding discrete model, compare Fig. \ref{fig:bending} and Fig. \ref{energy-1a}. In particular,  both models predict the abrupt emergence and subsequent proliferation of the non-affinity zones which contain inhomogeneously ruptured springs. Similarly, both models predict the abrupt recovery of the affine state as the rupture process saturates. 

Note  that the non-affine states, stabilized by bending (gradient) elasticity, can be characterized at the macro-scale in terms of   damage mechanics. It   provides a homogenized description of de-localized microcracking which  is relevant when the competing localized macro-cracking is inhibited. The  nonlocal stress redistribution, facilitated  by an under-constrained sub-system of stress-transmitting backbones, may be considered  a factor contributing to such inhibition.

\section{Elastic background} 
%The interplay between rigid and soft elements in such systems   leads to mechanical properties that can go far beyond the sum of those of the constituents and therefore  the   mechanical response of cross-linked fiber networks embedded in elastic matrix, has been addressed repeatedly, see for instance \cite{ L. Zhang, S. P. Lake, V. H. Barocas, M. S. Shephard, and R. C. Picu, Soft Matter 9, 6398 (2013).} ( \textbf{add reference}). 
To show that the de-localized damage can also appear in the form of periodic  \emph{patterns},  we now consider  our pantograph-reinforced  chain coupled to an elastic (Winkler's)  background. Breakable networks embedded in   soft elastic matrices are structural elements in many engineering materials and  biological materials, see for instance,  \cite{Van_Doorn2017-by, Cortes2012-be,Lynch2017-ol}. 
%Since in this paper we are targeting only the most basic aspects of the proposed meta-material model, we will continue using a quasi-one-dimensional setting, and we will leave the study of more realistic elastic environments for future research.

To achieve analytical transparency, we again assume   that the parameter $a$ is sufficiently small and adopt the quasi-continuum description encapsulated in \eqref{eq1}. Under the simplifying assumption that the elastic background is uniformly pre-stretched  with the same strain $\bar{\eps}$ as the composite chain, we can write the energy of the system in the form \cite{Vainchtein1999407,Ren:2000aa}  
\begin{equation}
\label{eq9}
E(u)= \int_0^1 \left( f(\eps)+\frac{\lambda_1^2}{2} \eps'^2 +\frac{\lambda_2^{-2}}{2}(u - u^0)^2 \right) dx,
\end{equation}
 where  $u^0(x) = (\bar{\eps}/2)(2x - 1)$ and $\lambda_2$ is a new dimensionless length scale characterizing the strength of the coupling; note that in \eqref{eq9} the elastic energy of the background is effectively subtracted. In what follows, we use the same boundary conditions on $u(x)$ as in the case of the unsupported chain. 
\begin{figure}[hbt!]
\centering
\includegraphics[scale=0.35 ]{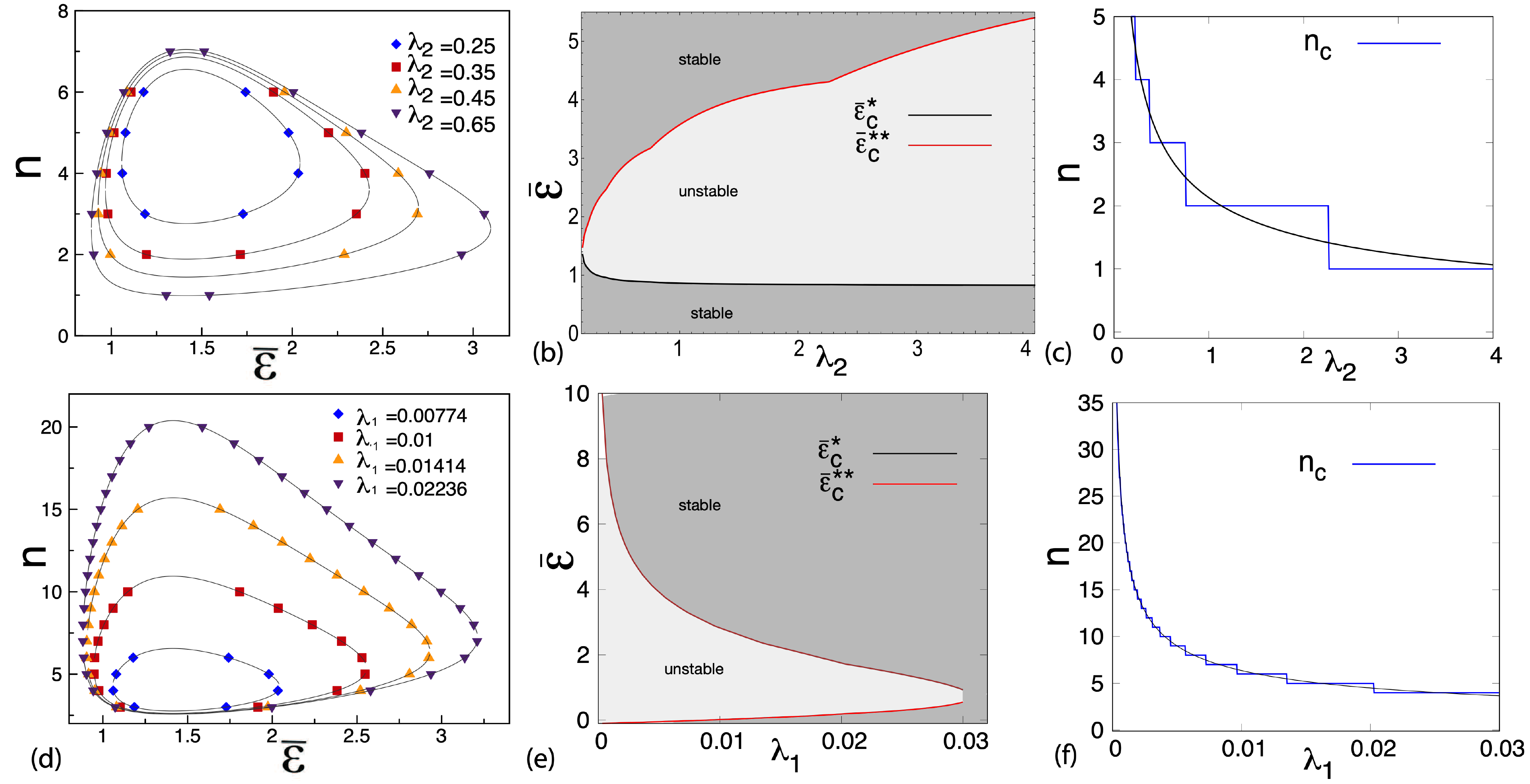}
\caption{Linear stability limits for the  homogeneous state in the continuum model of pantograph reinforced chain coupled to an elastic background: (a,d) bifurcation points, (b,e) critical strains  $(\bar{\eps}_c^*, \bar{\eps}_c^{**})$; (c,f) critical modes  $n_c$.  Parameters: in (a-c)  $\lambda_1 =0.02236$. In (d-f)  $\lambda_2 = 0.25$.  }
\label{fig2_3}
\end{figure}

In the model with energy \eqref{eq9}, the bifurcation  condition for the trivial branch can be found from the equation  generalizing   \eqref{eq3}
\begin{equation}
\label{eq111}
\lambda_1^2 (n\pi)^4 +  \frac{\de^2 f}{\de\eps^2} (\bar{\eps})(n\pi)^2 +  \lambda_2^{-2} = 0.
\end{equation}
 We  again obtain that there are upper and lower critical strains $\bar{\eps}_c^*$ and $\bar{\eps}_c^{**}$ corresponding to the same mode number $n_c$ so the bifurcation from the trivial branch is always of isola-center type.
 The difference from the case without the foundation is  that now one can have $n_c \neq 1$, see Fig.~\ref{fig2_3}.  To interpolate critical thresholds  we neglect the discreteness  and write  the approximating relations
$  \de^2 f/\de\eps^2({\bar{\eps}_c})= - 2 (\lambda_1/\lambda_2 )$ and  
 $n_c =  \pi^{-1} (\lambda_1\lambda_2 )^{-1/2}$. The expression for the critical wavenumber suggests that the instability is of Turing type \cite{Belintsev1981-wz,Toko1990-ao}.

 The dependence of solutions of \eqref{eq111} on parameters $\lambda_1$ and $\lambda_2$ is illustrated  in Fig.~\ref{fig2_3}(a,d); the parametric dependence of the corresponding bifurcation thresholds $\bar{\eps}_c^*$ and $\bar{\eps}_c^{**}$ is shown in Fig. \ref{fig2_3}(b,e). One can see that the non-affinity can be suppressed if the bending rigidity $\lambda_1$ is sufficiently large or if $\lambda_2$  and  the foundation coupling is stiff. However, the effect of the bending  in this respect is much stronger than the effect of the coupling, which needs to be infinitely stiff to eliminate non-affinity completely. Note also that the wave number of the unstable mode  tends to zero when either $\lambda_1$ or $\lambda_2$ disappears, see Fig.~\ref{fig2_3}(c,f). Interestingly,  the whole configuration of the  boundaries  in Fig.~\ref{fig2_3}(a,d) strongly resembles similar diagrams appearing in the fully nonlinear 3D theories of wrinkles in stretched elastic sheets \cite{Li2016-tq}.

\begin{figure}[hbt!]
\centering
\includegraphics[scale=0.32 ]{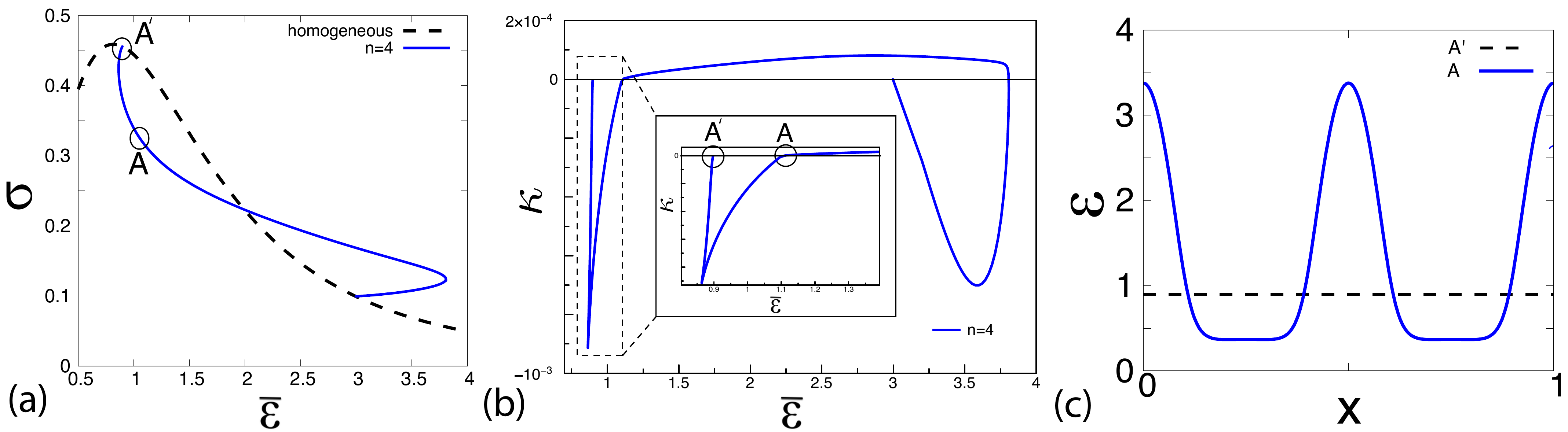}
\caption{Mechanical response in the  continuum model  of pantograph reinforced chain coupled to an elastic background:   (a) strain-stress relation along the first unstable branch $n=4$; (b) the smallest eigenvalue of the second variation for the same branch  as a function of the loading parameter $\bar\eps$; (c)  equilibrium strain profiles associated with the states A' and A. Parameters: $\lambda_1=0.0167$,  $\lambda_2=0.45$.}
\label{energy_0_1}
\end{figure}

The  study  of  the post-bifurcational behavior is based on   the nonlinear equilibrium equation 
\begin{equation}
\centering
\label{eq11}
-\lambda_1^2 u'''' +  \frac{\de^2 f}{\de\eps^2}(u') u'' -   \lambda_2^{-2}  (u - u^0) = 0.
\end{equation} 
 Eq. \eqref{eq11} describes the interplay between the localization tendency due to nonconvexity of the energy $f (u')$ and the delocalizing effect of weakly nonlocal bending elasticity brought by the pantograph reinforcement. This competition  is in turn affected by the bias towards homogeneity brought by  the interaction with an elastic foundation. The two   harmonic reinforcements can  be compared in terms of the effective interaction kernels \cite{Ren:2000aa}. Thus,  the pantograph sub-structure brings sign-definite, ferromagnetic-type interaction favoring coarsening of the damaged microstructure. Instead, the Winkler's foundation brings a sign indefinite,  anti-ferromagnetic type interaction favoring microstructure refinement. The complexity of the emerging post-bifurcation behavior reflects the general frustration due to  the presence of these competing tendencies.

Since the solution of \eqref{eq11} in quadratures is not available, we need to resort to numerical methods. Our computational approach is  the same as in the case  without   foundation and  is   based  on the use of the AUTO continuation algorithm \cite{doedel08auto-07p}.  The stability of  equilibrium branches is  again assessed by the numerical evaluation of the smallest eigenvalue of the second variation, which is now 
\begin{equation}
\delta^2 E(u)(v,v)=\int_0^1[(\de^2 f/\de\eps^2)(u')v'v'+\lambda_1^2v''v'' +  \lambda_2^{-2}vv]dx.
\label{hessian1}
\end{equation}
The corresponding stiffness matrix is 
%\begin{equation}
$K_{ij}=\int_0^1[(\de^2 f /\de\eps^2)(u'){\mathcal N}'_i {\mathcal N}'_j+\lambda_1^2 {\mathcal N}''_i {\mathcal N}''_j  + \lambda_2^{-1}{\mathcal N}_i{\mathcal N}_j]dx.$
%\label{hessian11_stifness_subs}
%\end{equation} 

Our first illustration concerns  the  nucleation event at $\bar{\eps}=\bar{\eps}_c^*$. The    bifurcated equilibrium branch,  corresponding to  $n=4$, is shown in Fig.~\ref{energy_0_1}(a). One may expect   the instability at the point A'  to result  in the  transition A'$\to$A where the non-affine configuration A also lies on the $n=4$, see Fig.~\ref{energy_0_1}(c). However,   the study of the smallest eigenvalue of the second variation \eqref{hessian1}  for  the equilibrium branch $n=4$shows that the corresponding  state A at $\bar{\eps}=\bar{\eps}_c^*$ is unstable, see  Fig.~\ref{energy_0_1}(b). According to this figure there is a finite gap separating the first point of instability of the homogeneous state and the  first stable equilibrium  with $n=4$, see point A in  Fig.~\ref{energy_0_1}(a,b). 
 \begin{figure}[hbt!]
\centering
\includegraphics[scale=0.4 ]{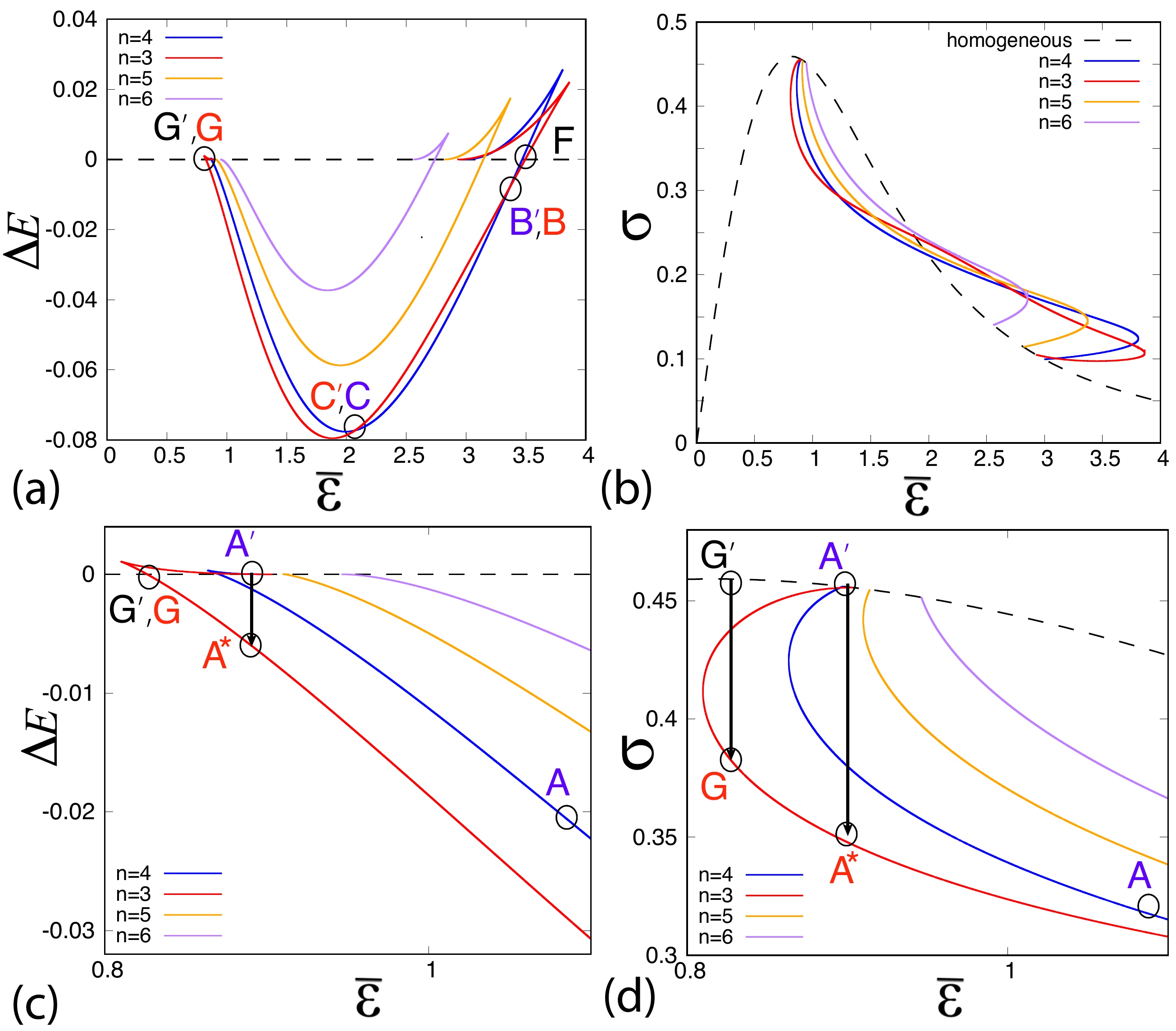}
\caption{Mechanical response in the continuum model of a pantograph reinforced chain coupled to an elastic foundation:   (a) the energy difference between the   non-affine  and affine equilibrium configurations; (b) macroscopic strain-stress relations along different equilibrium branches; (c,d) magnified versions of (a) and (b) near the first instability point. The transition 
 A'$\to$A$^*$ takes place along the LEM path while  the transitions  G'$\to$G,C'$\to$C, B'$\to$B and F$\to$F' are the equilibrium branch switching events taking place along the GEM path. Parameters: $\lambda_1=0.0167$,  $\lambda_2=0.45$.}
\label{energy_0}
\end{figure}

The presence of   competing interactions in this problem suggests that the stable pattern, emerging from the decomposition of the homogeneous state, may be \emph{different} from the one implied by the linear stability analysis. The global picture is presented in Fig. \ref{energy_0}(a,c) where we show the energy-strain and the stress-strain relations along  the equilibrium branches with $n=3,..., 6$. They all bifurcate from the homogeneous branch around $ \bar{\eps}_c^*=0.897334$, for instance the second bifurcation point (after the one with $n=4$) at   $ \bar{\eps}_c=0.903472$ corresponds  to the branch with $n=3$. 

\begin{figure}[h!]
\centering
\includegraphics[scale=0.6 ]{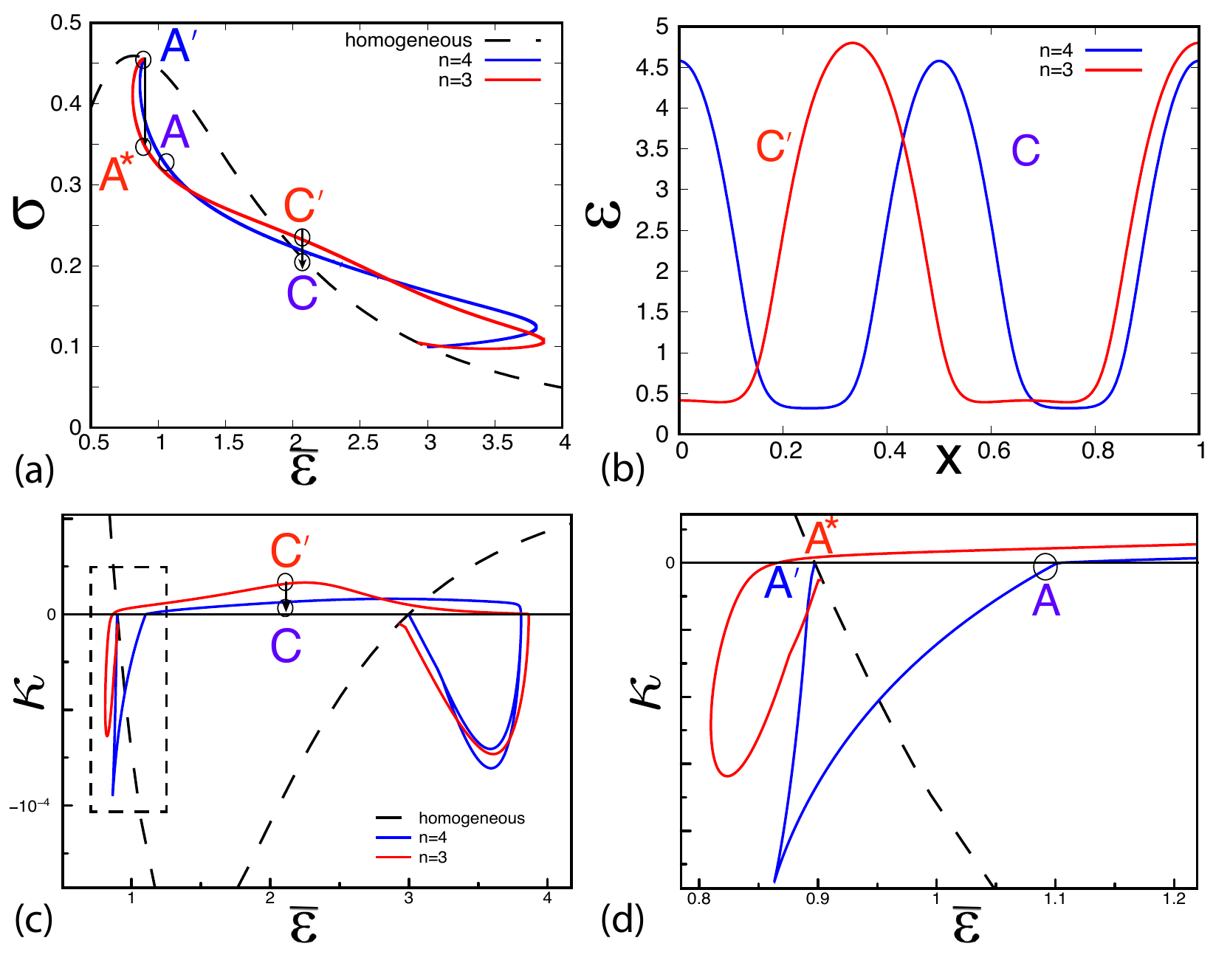}
\caption{ Zoom in on the particular branch switching events along LEM and GEM paths shown in Fig.\ref{energy_0}: (a) stress-strain response, (b) equilibrium configurations C and C' with the same energy, (c) smallest eigenvalue of the second variation for the branches with $n=3$ and $n=4$, (d) zoom in on (c) around the first instability along the LEM path. Symmetry is lost during the transition  A' $\to$ A$^*$ while  symmetry is acquired during the C'$\to $C transition. Parameters: $\lambda_1=0.0167$,  $\lambda_2=0.45$.}
\label{globalminmech}
\end{figure}

The detailed picture is shown in Fig. \ref{energy_0}(b,d) where we see that, in view of the instability of the $n=4$ branch in the interval of interest, the only   configuration reachable from point A' by energy minimization  is the one corresponding to point  A$^*$ on the $n=3$ branch. The local stability of this branch  is illustrated  in the inset in Fig.~\ref{globalminmech}(c). Moreover, in this range of strains, the equilibrium branch with $n=3$    also delivers  the \emph{global} minimum of the energy.

\begin{figure}[h!]
\centering
\includegraphics[scale=0.6 ]{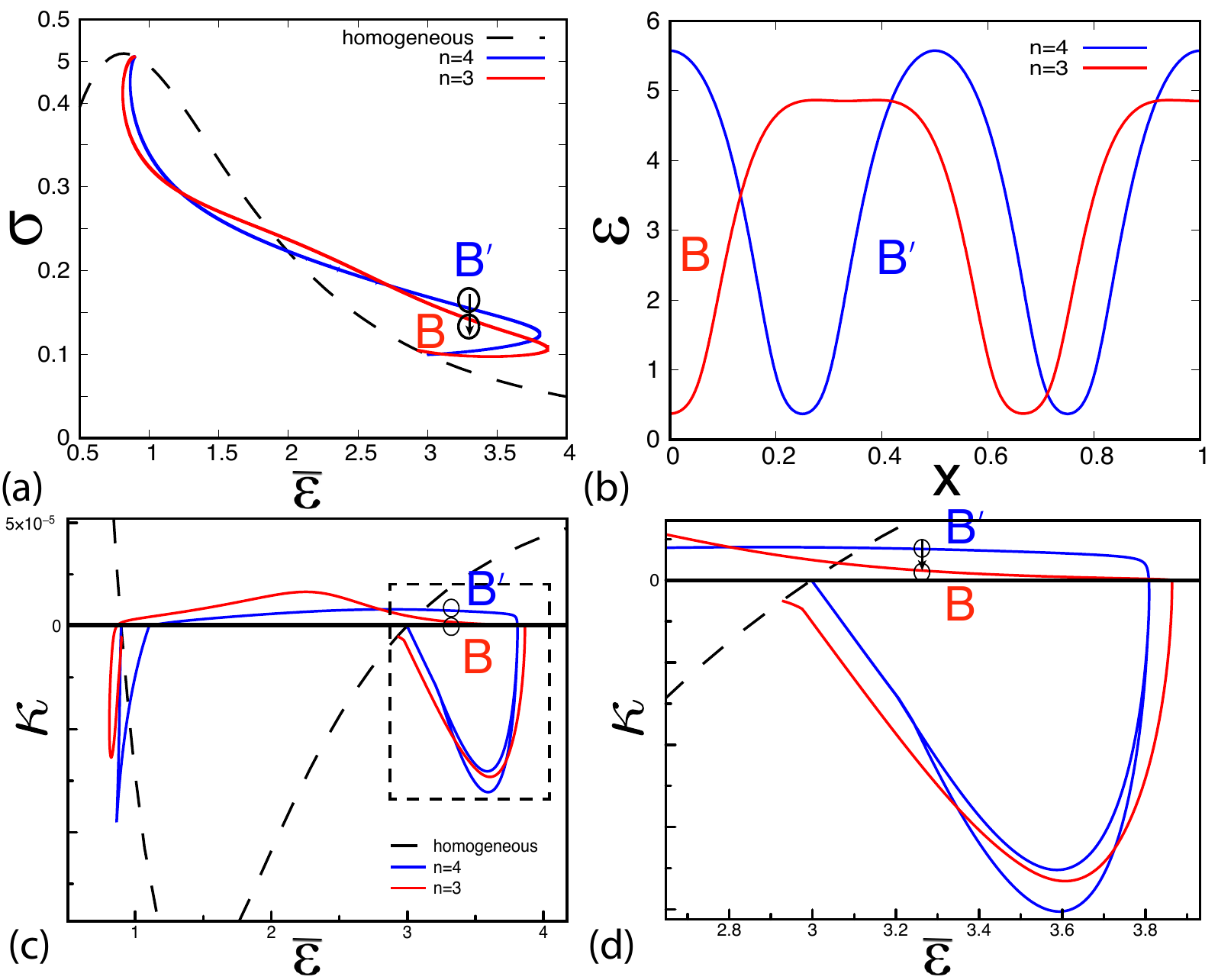}
\caption{Zoom in on the particular branch switching events along LEM and GEM paths shown in Fig.~\ref{energy_0}: (a) stress-strain response, (b) equilibrium configurations B and B' with the same energy, (c) smallest eigenvalue of the second variation for the branches with $n=3$ and $n=4$, (d) zoom in on (c) around the  transition B'$\to $B  along the GEM path.   Symmetry is lost during the B'$\to$B transition. Parameters: $\lambda_1=0.0167$,  $\lambda_2=0.45$.}
\label{globalminmech1}
\end{figure}

Considering   GEM dynamics, we can conclude  that the transition from the trivial branch $n=0$, taking place at the  point G (Fig. \ref{energy_0}), also leads to the branch with $n=3$. According to Fig.~\ref{energy_0}(a,c)  the next transition along   the  GEM  path takes place at point C' and  brings the system back to $n=4$ equilibrium branch. Then, at point B' the system returns again on the $n=3$ branch and finally   stabilizes  on the trivial $n=0$ branch at point F. In Fig.~\ref{globalminmech}(b) and Fig.~\ref{globalminmech1}(b)  we provide evidence  that 
these GEM transitions are between locally stable states. 

Our Fig.~\ref{globalminmech}(c) and Fig.~\ref{globalminmech}(c) illustrate the nature of the restructuring of the equilibrium configurations during these transitions. Thus,
as  the   loading parameter $\bar{\eps}$ increases, the  non-symmetric configuration produced at point G, which contains  two domains of non-affinity and corresponds  to the branch with $n=3$,   evolves   till  another  strain threshold is reached  where the system undergoes the  transformation  C'$\to$ C   to the  symmetric configuration  corresponding to $n=4$. As a result, the second  surface-bound damage zone appears, see Fig.~\ref{globalminmech}(c). Then,  during the  reverse transition B'$\to$B   the symmetry is lost  and  one of the     surface-bound   damage zone   disappears again, see  Fig.~\ref{globalminmech1}(c).  Finally, the symmetry is recovered when the affine configuration  stabilizes at point  F. 

In contrast to the complexity of the GEM path, the LEM dynamics produces much simpler set of transitions. Without going into details, we only mention that after the dissipative A$\to$A$^*$ transition the system remains on the $n=3$ branch all the way till its reaches the turning point where another dissipative  transition to the homogeneous branch $n=0$ takes place (not shown in either of the figures).

Note that the parameters in  Fig. \ref{energy_0}, Fig.~\ref{globalminmech}  and Fig.~\ref{globalminmech1}  were chosen arbitrarily with the only consideration  that the number of the domains of nonaffinity is relatively small. It is straightforward to see that the microcracking pattern in this problem may be arbitrarily complex, for instance in the limit $\lambda_1 \to 0$ the number of cracks tends to infinity. In fact, by choosing the bending resistance of the metamaterial sub-structure and  the degree of coupling with the elastic environment, one can  effectively control the complexity of the  emerging microstructures.

More generally,  the above analysis shows that, despite the constituents' brittle nature, the material response of the pantograph-reinforced chain is incompatible with the conventional scenario of highly localized cracking.
 Instead, the  model predicts the emergence of  de-localized damage zones, which, in the presence of elastic background,  advance from multiple sources and form regular patterns. Competitive interactions ensure that a monotonously loaded system  experiences a series of instabilities where symmetries may be lost and re-acquired.
The interplay between two internal length scales in this problem may be a source of the considerable complexity in the spatial distribution of damage.

\section{Conventional brittle fracture} 

Since one  internal   length scale is also present  in the conventional fracture mechanics and  another one can be  added through the coupling
 to an elastic foundation, one may ask if the pantograph-based sub-structure brings anything fundamentally different. To answer this question we consider  in this Section a continuum model of fracture where cracks are described  by a phase-field~\cite{PhysRevLett.87.045501,Lorentz2003-oj,Pham20111163}. 

In  phase-field  theories  of fracture,  elasticity of the breaking  solid is  usually assumed to be  linear  with  stiffness  degrading with damage. The latter is  described by a scalar  order parameter  with the square of the gradient of this   parameter  controlling the energy cost of the the  damage non-affinity~\cite{Barenblatt1993-uu}.  

Suppose that the   damage  variable is  $\a(x)$  with  $\a=0$ ($\a=1$) corresponding to the unbroken (fully broken) state.  In the absence of  the elastic environment,  we can  write  the   energy of the system  in the form
\begin{equation}
\label{eq5}
E(u,\alpha)=\int_{0} ^1 \left( f(\eps,\alpha)
+\frac{\lambda_1^2}{2}(\a')^2 \right)dx.
\end{equation}
where   again $\eps(x)=u'(x)$. We assume that the local energy density in \eqref{eq5}  is of the form
\begin{equation}
\label{eq55}
f(\eps,\alpha)=\frac{1}{2}g(\a)\eps^2 + h(\a),
\end{equation}
where  the first term on the right, which is quadratic in strain $\eps(x)$,  describes  linear elasticity at constant damage. The second term,  depending only on $\a(x)$,  is the  energetic   price  of   homogeneous   damage.  The gradient term in \eqref{eq5}   penalizes the  inhomogeneity of the damage  and brings into the theory an internal length scale $\lambda_1$. 
The  equilibrium branches  in this model are represented by solutions   of the nonlinear  equations
\begin{eqnarray}\label{modeld_el}
\begin{cases}
  -(g(\a)u')' &= 0 \\
  -\lambda_1^2\a'' + \frac{1}{2} \frac{\partial g}{\partial \a}(u')^2+ \frac{\partial h}{\partial \a} &= 0
\end{cases}
\end{eqnarray}

To mimic  in this framework our numerical   experiments with the breakable  chain, see Fig.~\ref{fig1a},  we  make the standard assumptions that  
$g(\a)= (1-\a)^2$, and $ h(\a)=\a^2$. if we also   choose the  boundary conditions  in the form  $u(0)=-\bar \eps/2$, $u(1)= \bar \eps/2$ and   $\a'(0) = \a'(1) = 0,$ the   homogeneous solution, representing in this case the principal branch of equilibria,  takes the form
\begin{equation}\label{eq:stability1}
u^0(x) =  (\bar\eps/2)(2x-1),\,\, \a^0(x) =  \bar\eps^2/(2+\bar\eps^2).
\end{equation}
The effective elastic energy along the trivial branch is 
\begin{equation}\label{eq:stability1}
f^0(\eps) =f(\eps, \a^o(\eps)) =  \eps^2/(2+\eps^2).  
\end{equation}
whose similarity with potential adopted in the study of our Ginzburg-Landau elastic model justifies the  assumptions for $g(\a)$ and  $ h(\a)$.  We may therefore perceive  the   phase-field model as a version of an elastic  theory  with the softening energy $f^0(\eps) $  \cite{Pellegrini:2010aa}.

\begin{figure}[h!]
\begin{center}
   \includegraphics[scale=.5]{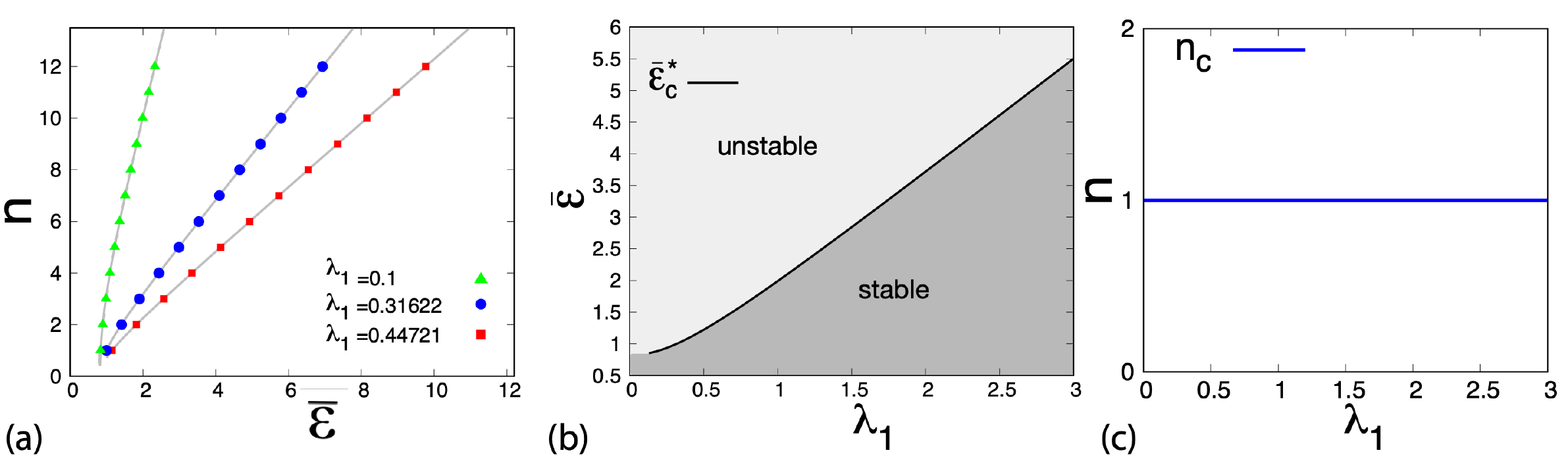}
  \caption{Stability study of the homogeneous (affine) state in the phase-field model of a unreinforced breakable chain: (a) stability boundaries (bifurcation points)  depending on the regularization parameter $\lambda_1$; (b) critical strains $\bar\eps_c$; (c) critical mode $n_c$.}
  \label{fig:fig10}
  \end{center}
\end{figure}

To test the linear stability of the homogeneous solution \eqref{eq:stability1} we linearize \eqref{modeld_el} and look for nontrivial solutions  $s(x)=u(x)-u^0(x)$ and $\tau(x)=\a(x) -\a^0$.  Following the same procedure as in the previous sections, we obtain that the unstable modes are  $s(x) \sim \sin(n\pi x)$  and $\tau(x)\sim \cos(n\pi x)$  with the bifurcation condition taking the form
\begin{equation}\label{eq:stability}
(n\pi)^2 = - \frac{4\bar\eps^2  \de^2 f^0/\de\eps^2(\bar\eps)}{\lambda_1^2[(\de f^0/\de\eps(\bar\eps)) \bar\eps -   \de^2 f^0/\de\eps^2(\bar\eps)]}.
\end{equation}
In Fig.~\ref{fig:fig10}(a)  we illustrate the $\lambda_1$ dependence of the solutions of \eqref{eq:stability}.   
% (1/2)g(A(\eps))\eps^2 + h(A(\eps)) 
As in the case of pantograph-reinforced chain, the critical mode is always   $n_{c} = 1$, see   Fig.~\ref{fig:fig10}(c).

To see the difference, we now consider   the $\lambda_1$ dependence of   the critical strain $\bar\eps_{c}$ which can be found from the equation 
\begin{equation}\label{crit-damage-k=0}
\bar\eps_c(4\bar\eps_c^2 - \lambda_1^2\pi^2)\frac{   \de^2 f^0/\de\eps^2(\bar\eps_c)}{  \de f^0/\de\eps(\bar\eps_c)} + \lambda_1^2\pi^2 = 0.
\end{equation}
It takes particularly simple form  for our choice of the potential $f^0(\eps)$   when we can write
$ \lambda_1 ^2 \pi^2= 3  \bar{\eps}_c^2-2$, see Fig.~\ref{fig:fig10}(b).
%It can be desumed that, in order to destabilize the homogeneous state, $\frac{\de^2\phi(\eps)}{\de\phi(\eps)}<0$ as also previously found.
%Note, but, this case has a bifurcation criterion rather different from \eqref{bifurcation_without} since it also involves the strain $\eps$ alone.
%This would imply that a result analogous to the one in Prop. \eqref{prop-stabregion} cannot be found.
%This would have relevant consequences, that will be shown in an example.
%For the choice of potential $\phi(\eps) = \frac{\eps^2}{2+\eps^2}$ this condition \eqref{crit-damage-k=0} thus, as can be easily seen, admits an unique solution $\eps_{crit}$.
Now,   in contrast to the case of the  pantograph-reinforced  chain,  see  Fig.~\ref{fig1b}, the affine  configuration  does not re-stabilizes  after the initial  instability and  the 'broken' configuration always remains  non-affine, see Fig.~\ref{fig:fig10}(b). Such response is in full agreement with the behavior of  a simple breakable chain, see  Fig. \ref{fig1a}, but despite the presence of gradient term in \eqref{eq5}, the effect of the pantograph reinforcement is lost.

 %  for the  general  potential growing sub-linearly like $\eps^{p}$ ($p<1$).
%If this is the case, $\eps(4\eps^2 - \lambda_1^2)\frac{\de^2\phi(\eps)}{\de\phi(\eps)} \sim -\eps^{2}$ for large $\eps$.
%Then equation \eqref{crit-damage-k=0} for large strain and small $\lambda_1^2$, becomes $ \eps^{2} = \lambda_1^2$ which has no solutions for $\eps>\sqrt{\lambda_1^2}$.
\begin{figure}[h!]
\begin{center}
   \includegraphics[scale=.3 ]{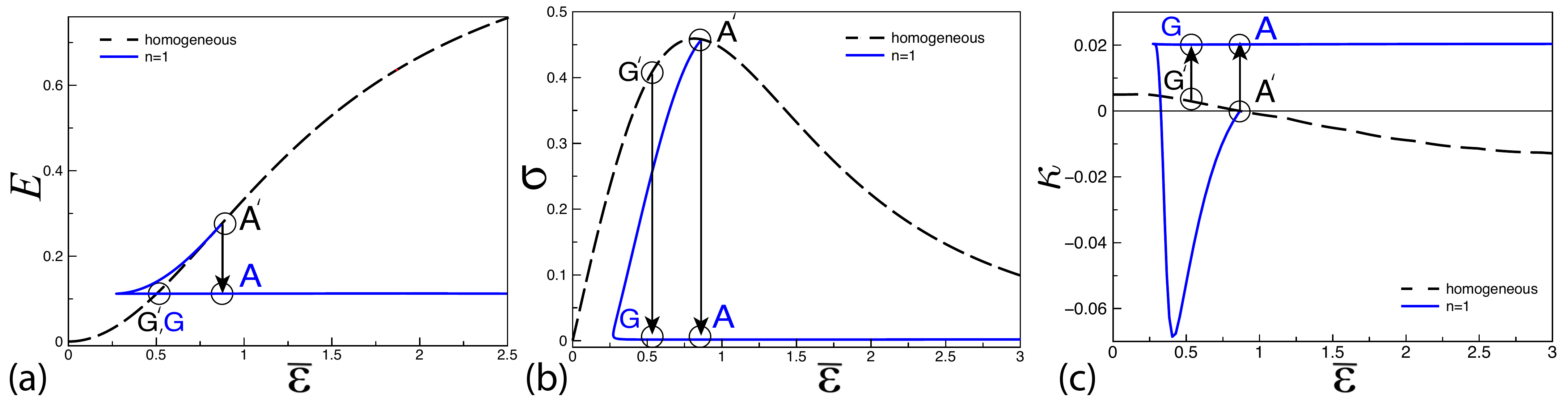}
  \caption{Mechanical response in the  phase-field  model of unreinforced chain along the LEM (A' $\to$A transition) and the GEM (G' $\to$G transition)  loading paths: (a) macroscopic energy-strain relation; (b) macroscopic stress-strain relation; (c) smallest eigenvalue of the second variation for the homogeneous branch  as a function of the loading parameter $\eps$.   Single macroscopic crack nucleation takes the form of   A'$\to$A transition.
  Here, $\lambda_1=0.158114$.}
  \label{pf_nl}
  \end{center}
\end{figure}
The  fracture localization (vs de-localization) in the phase-field model  is illustrated in   Fig.~\ref{pf_nl}.   We combine there  the  solutions of the nonlinear system  \eqref{modeld_el} corresponding to   LEM and   GEM protocols.  To confirm   the local stability of  the  configurations found by  our LEM  algorithm, we computed  the lowest eigenvalue of the second variation 
%\begin{equation}
%\delta^2 E(u,\alpha)(v,w)=\int_0^1 [g(\alpha)v'v' +\frac{1}{\lambda_2^2}vv+ 2g'(\alpha)u' v'w+(h''(\alpha)+\frac{u'^2}{2}g''(\alpha))ww+\lambda_1^2w'w']dx,\label{hessian2}\end{equation}
 \begin{equation}
\delta^2 E(u,\alpha)(v,w)=\int_0^1 [(1-\alpha)^2v'^2 
%+ \lambda_2^{-2}v^2
- 4(1-\alpha)u' v'w+(2+ u'^2)w^2+\lambda_1^2w'^2]dx,\label{hessian2}\end{equation}
where $v$ and $w$ are  test functions. The stiffness matrix  is then 
 %${\bf K}$
% in the form 
\begin{equation}
{\bf K}=\begin{bmatrix}
\int_0^1[(1-\alpha)^2{\mathcal N}'_i{\mathcal N}'_j
%+\frac{1}{\lambda_2^2}{\mathcal N}_i {\mathcal N}_j
] dx&
-2\int_0^1(1-\alpha)u' {\mathcal N}'_i {\mathcal N}_j  dx\\
-2\int_0^1(1-\alpha)u' {\mathcal N}_i {\mathcal N}'_j dx
& \int_0^1 [ (2+u'^2){\mathcal N}_i{\mathcal N}_j +\lambda_1^2{\mathcal N}'_i{\mathcal N}'_j] dx
\end{bmatrix},
\end{equation}
where the shape functions ${\mathcal N}_i(x)$  can be  now  chosen  simply quadratic.

According to  Fig.~\ref{pf_nl}, the LEM dynamics, which reduces  to local energy minimization,   is characterized by a  major dissipative event in the form of the transition  A' $\to$A from the branch $n=0$ to the branch $n=1$.  Instead,  the GEM dynamics, which implies   global energy minimization,  is epitomized by an earlier and smaller  non-dissipative transition G' $\to$G, also from the branch $n=0$ to the branch $n=1$.  Note that  the metastable  section  of the non-affine equilibrium branch $n=1$ in Fig.~\ref{pf_nl} was constructed  using  again the pseudo-arclength continuation technique  implemented in the software package AUTO   \cite{doedel08auto-07p}. 

\begin{figure}[h!]
\begin{center}
   \includegraphics[scale=.3 ]{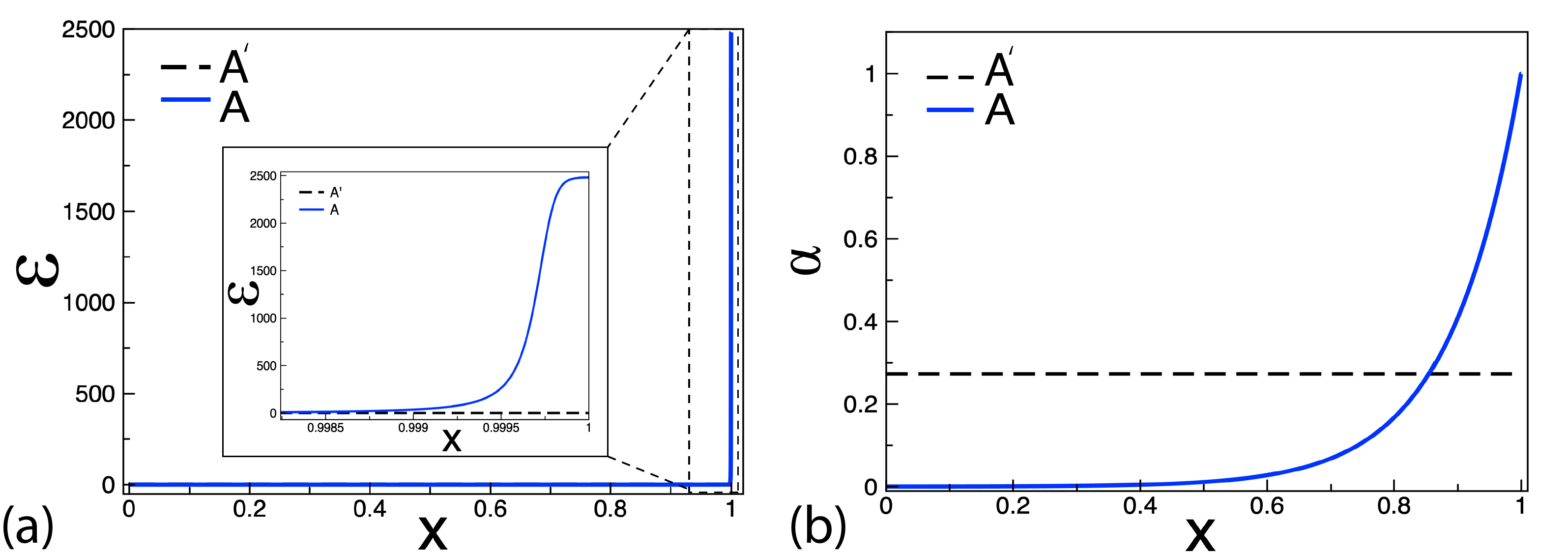}
  \caption{The detailed picture of the  A'$\to$A transition shown in Fig.  \ref{pf_nl}. The macroscopic crack is nucleated near the boundary of the sample: (a) strain profile, (b) damage parameter profile (b). Here, $\lambda_1=0.158114$.}
  \label{pf_nl1}
  \end{center}
\end{figure}

We observe, see   Fig.~\ref{pf_nl}(a,b), that  the  stretching  response of this continuum system is basically the same as for the simple chain with breakable elements. The fact that the crack forms on one of the   boundaries is due to a small bias provided by the phase-field  related  boundary conditions.  After the major stress drop the subsequent loading does not create additional damage and the response reduces  to the  increase  of the amplitude  of the localized strain. Note again that during the stress drop, the system does not fully unload because some  (weakening) elastic interaction between the newly formed crack lips always exists. The noteworthy difference between the behavior of the original discrete system and its   phase-field analog  is the   smearing out of the crack  due to the gradient regularization, see Fig.~\ref{pf_nl1}(a,b). If the strain remains sufficiently localized, the damage parameter shows an extended  boundary layer whose  structure, however,   is fundamentally different from the diffuse damage configuration  generated  in  the pantograph-reinforced chain. 
%Due to this difference, the  phase-field regularization is ultimately unable to    recover the homogeneous state at large levels of stretching.

Our example shows that the phase-field  framework cannot be used alone to build a continuum model of the pantograph-reinforced structure. In particular, we show that due to weaker regularization through $\a$ vs regularization through $\eps$, the standard phase-field model does not capture the reentry nature of the bifurcation   and therefore misses the main effect: the recovery of the homogeneous state at a large levels of stretching. We have seen that in phase-field  theory, fracture remains localized independently of the loading, and  the phenomenon of diffuse  microcracking does not take place. 

Below we show that the introduction of  an elastic foundation in the phase-field framework fixes the problem only partially.  
 In this case, the return to the affine configuration  at large levels of stretch is secured due to the ultimate dominance of the elastic foundation, however,  the task of recovering the diffuse microcracking remains elusive.

Indeed,  consider the energy functional of the form
\begin{align}\label{modeld}
     E = \int_{0}^{1} \left(f(\eps,\a)
     %\frac{1}{2} g(\a)(u')^2 + h(\a)
      + \frac{\lambda_1^2}{2} (\a')^2 +  \frac{\lambda_2^{-2}}{2} (u - u^0)^2 \right)dx,
\end{align}
 where again
$u^0(x) =  (\bar\eps/2)(2x-1)$. We keep the same boundary conditions as in the case without elastic foundation.  Then again the   equilibrium equations   
\begin{eqnarray}\label{modeld_elf}
\begin{cases}
  -(g(\a)u')' +  \lambda_2^{-2}(u-u^0) &= 0 \\
  -\lambda_1^2\a'' + \frac{1}{2}(u')^2  \frac{\partial g}{\partial \a}+  \frac{\partial h}{\partial \a} &= 0.
\end{cases}
\end{eqnarray}
%have  a  trivial solution of the form $u^0(x), \alpha^0(x)$.  
We can now apply the same computational approach as in the  foundation free case while appropriately modifying the expression for 
% case,  dealing with two unknown functions $u(x)$ and $\a(x)$.  To study stability, we use 
  the second variation 
%\begin{equation}
$\delta^2 E(u,\alpha)(v,w)=\int_0^1 [(1-\alpha)^2v'^2 
 + \lambda_2^{-2}v^2
- 4(1-\alpha)u' v'w+(2+ u'^2)w^2+\lambda_1^2w'^2]dx.$

The wavenumber of  nontrivial perturbations can be found from the   equation   
\begin{equation}\label{bifurcation_d}
  \lambda_1^2 \frac{g(\a^o)}{\frac{\eps^2}{2}\frac{\de^2 g}{\de \alpha^2}(\a^o) + \frac{\de^2 h}{\de \alpha^2}(\a^o)} (n\pi)^4  +  \left( g(\a^o) + \frac{ (\frac{\lambda_1}{\lambda_2})^2  -(\eps \frac{\de g}{\de \alpha}(\a^o))^2}{\frac{\eps^2}{2}\frac{\de^2 g}{\de \alpha^2}(\a^o) + \frac{\de^2 h}{\de \alpha^2}(\a^o)} \right)(n\pi)^2 + \lambda_2^{-2}= 0,
\end{equation}
\begin{figure}[h!]
\centering
   \includegraphics[scale=0.4]{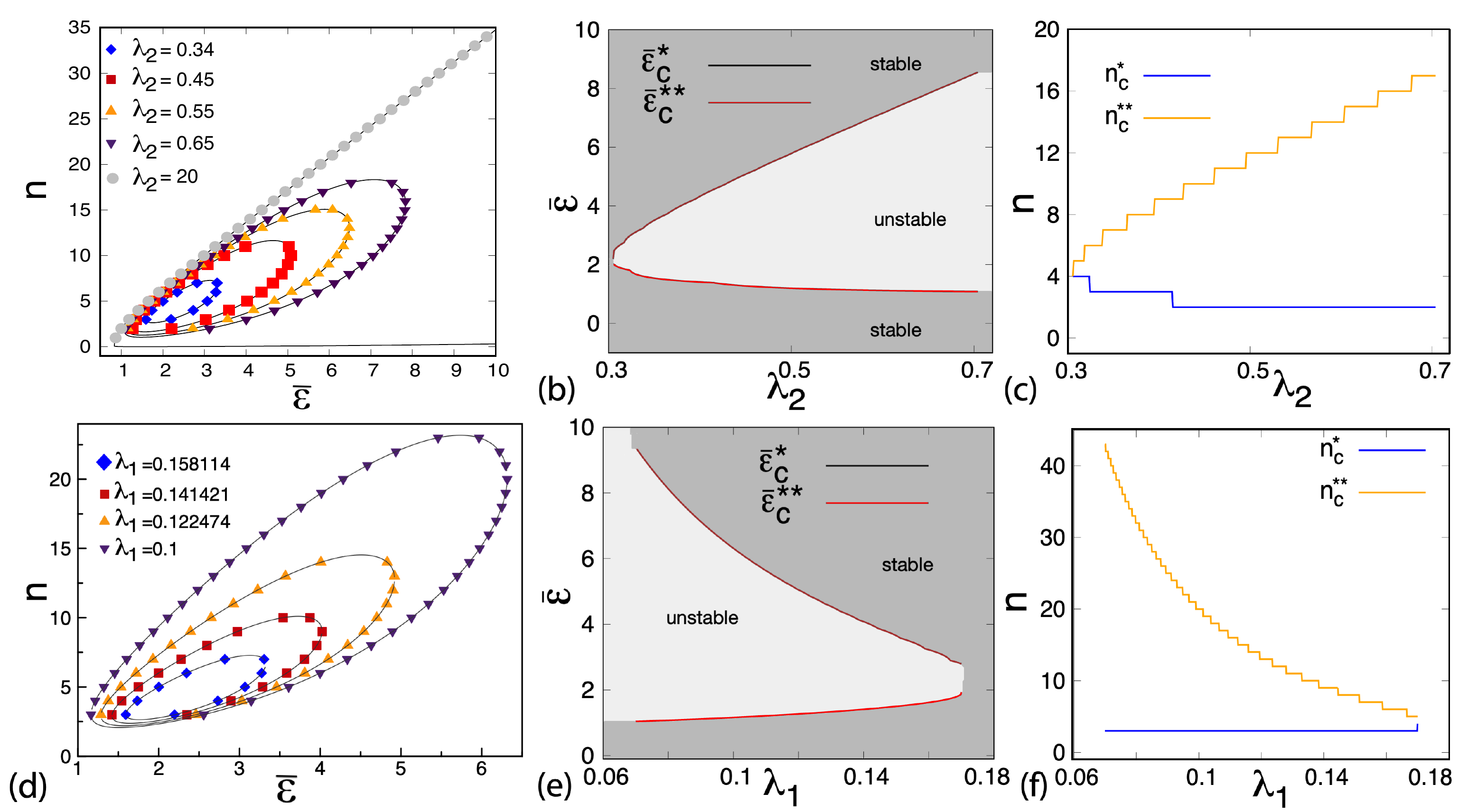}
  \caption{Linear stability of a homogeneous state in the continuum (phase-field)  model  of breakable chain coupled to an elastic background. Parametric study:  (a,d) bifurcation points; (b,e) critical strains $\bar{\eps}_c^*$ and $ \bar{\eps}_c^{**}$; (c,f)   instability mode numbers  $ n_c^*$ and $n_c^{**}$.  In (a-c)  $\lambda_1=0.158114$ and in (d-f)  $\lambda_2=0.34$.}
  \label{energy-ex01}
\end{figure}
In terms of the   effective elastic energy density $f^0(\eps)$
%=\eps^2/(2+\eps^2)$ 
 the  bifurcation condition for $\bar\eps_{c}$ reads
\begin{equation}
\label{eq_n2}
 \left[ \frac{\lambda_1^2}{\lambda_2^2}\frac{ f^0 (\bar\eps_{c})}{\bar\eps_{c}^2} +  \frac{\de^2 f^0}{\de\eps^2} (\bar\eps_c)\right] ^2 - \frac{\lambda_1^2}{\lambda_2^2}\frac{1}{\bar\eps_{c}^2}\left[ \frac{ 1}{\bar\eps_{c}} \frac{ \de f^0}{\de\eps}(\bar\eps_c)-    \frac{\de^2 f^0}{\de\eps^2} (\bar\eps_c)\right]  = 0.
\end{equation}
Then the   critical mode number $n_c$  is
\begin{equation} 
\label{eq_n}
(n_c\pi)^2 = -\frac{ (\frac{\lambda_1}{\lambda_2})^2\frac{f^0 (\bar\eps_{c})}{\bar\eps_{c}^2} +     \frac{\de^2 f^0}{\de\eps^2} (\bar\eps_c)}{\frac{\lambda_1^2}{2\bar\eps_{c}^2}\left[ \frac{ 1}{\bar\eps_{c}} \frac{ \de f^0}{\de\eps}(\bar\eps_c)-    \frac{\de^2 f^0}{\de\eps^2} (\bar\eps_c)\right]} .
\end{equation}
The parametric dependence of the  solutions of \eqref{bifurcation_d},  is illustrated in Fig.  \ref{energy-ex01}(a,d). We see the return of the closed loops  as in the case of a pantograph-reinforced chain, see Fig. \ref{fig2_3}(a,d). 
%This implies  that we are again confronting the  isola-center bifurcation. 
The  loops becomes larger as $\lambda_2 \to \infty$, and in the limit,  we recover the  loopless case of   classical brittle fracture, see  Fig. \ref{fig:fig10}(a).  Such `opening'  of the stability boundaries is reminiscent of  the theory of wrinkles in stretched elastic sheets   as one moves from the fully nonlinear 3D  theory for, say neo-Hookean material, to a simplified   Foppl-von Karman theory~\cite{Li2016-tq}.

In Fig.  \ref{energy-ex01}(b,e), we show the parametric dependence of the critical strain, see   \eqref{eq_n2}. Due to the presence of an elastic background, the re-entry behavior of the affine configuration is recovered with the emergence  of the two critical strains $ \bar\eps_{c}^*$ and $\bar\eps_{c}^{**}$ representing, respectively, the lower and upper limits of stability for the homogeneous state, see Fig.  \ref{energy-ex01} (b,e).
% The re-entry feature  disappear as the elastic coupling with the foundation weakens and in the limit  $\lambda_2 \to \infty$ become indistinguishable, see Fig.  \ref{energy-ex01}(a) and Fig. \ref{fig:fig10}(a).
The parametric dependence of the critical wavenumber $n_c$,  shown in Fig. \ref{energy-ex01}(c,f), 
%The obtained behavior 
departs from what we have seen in the model of   pantograph-reinforced chain on an elastic foundation as the critical wavenumber $n_c(\bar{\eps}_c^*)$ is now different from the critical wavenumber $n_c(\bar{\eps}_c^{**})$. We can link this result with the fact that the re-stabilization (or healing) of the affine state at large levels of stretching is enforced by a different physical mechanism in our two settings: the bending induced `weak' nonlocality of `ferromagnetic' type in the case of the pantograph-reinforced breakable chain, and the elastic foundation-induced `strong' nonlocality of anti-ferromagnetic type in the case of the chain coupled to an elastic foundation. 

\begin{figure}[hbt!]
\centering
\includegraphics[scale=.251]{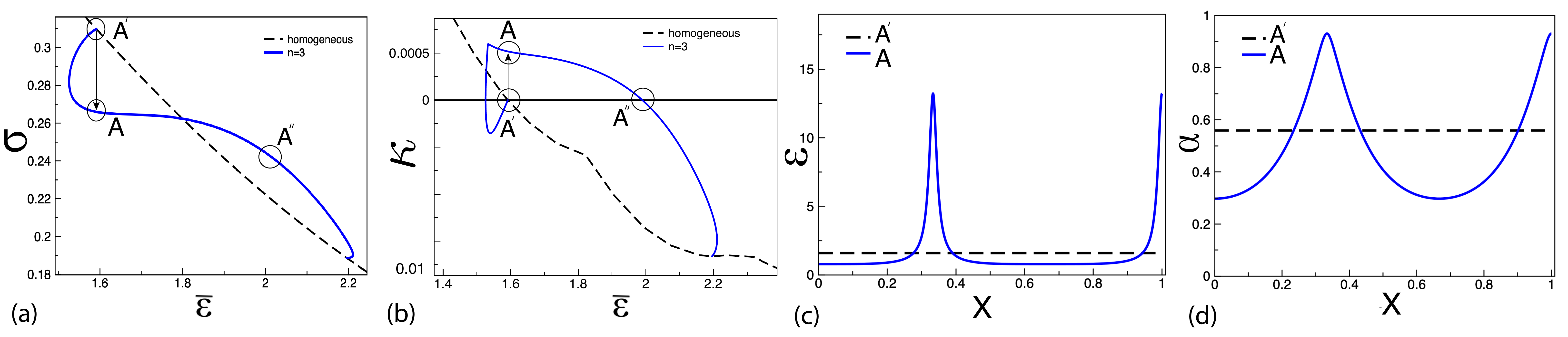}
\caption{First instability along the LEM path in the continuum model of a breakable chain coupled to an elastic background:  (a) macroscopic strain-stress response showing the discontinuous transition A'$\to$A  from affine to non-affine states  at the critical strain $\bar{\eps}_c^{*}$; (b) smallest eigenvalue of the second variation for the  branch with $n=0$ and $n=3$ as a function of the loading parameter $\bar\eps$; (c,d) strain   and damage parameter  profiles  before and after the A'$\to$A transition. Parameters: $\lambda_1=0.158114$, $\lambda_2=0.34$.}
\label{fig2_4_2}
\end{figure}

Consider now the post-bifurcational response under  the LEM protocol.
\begin{figure}[h!]
\centering
\includegraphics[scale=.3]{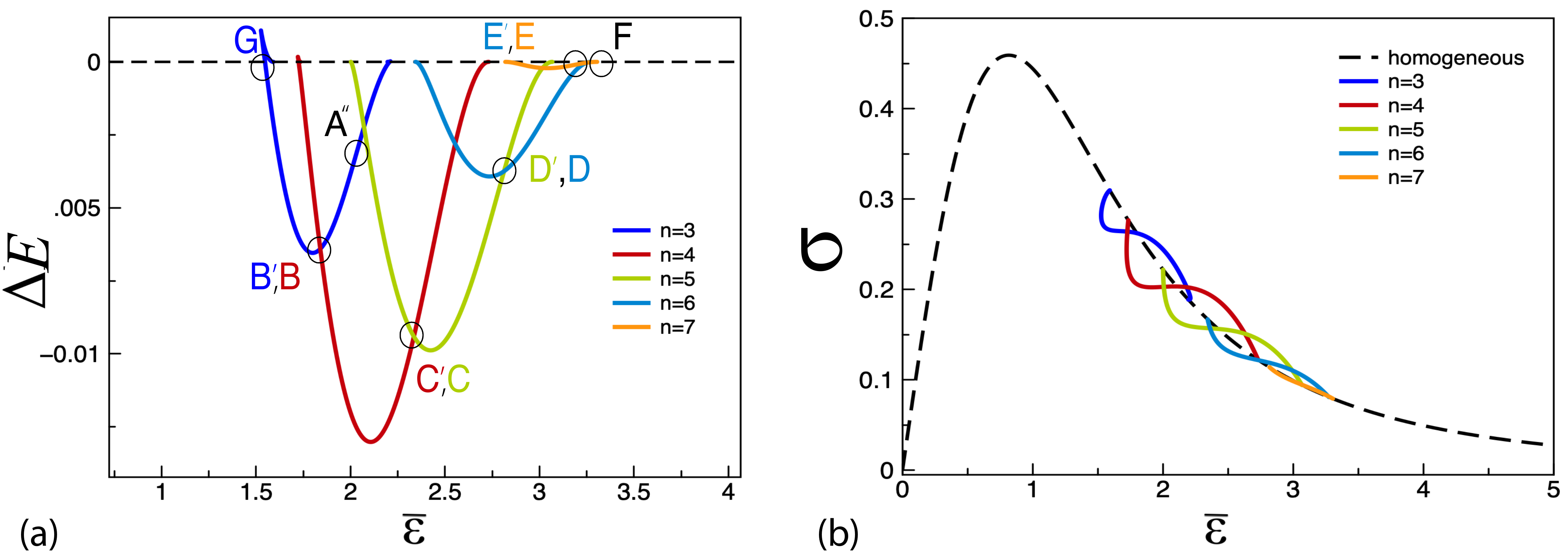}
\caption{Mechanical response in the continuum (phase-field) model of a breakable chain on elastic foundation deformed according to the GEM protocol: (a) the energy difference between the affine and   non-affine  configurations, letters G,B,C,D,E,F indicate dissipation-free  branch switching events; (b) the corresponding macroscopic strain-stress relations. Parameters: $\lambda_1=0.158114$, $\lambda_2=0.34$.}
\label{fig2_4}
\end{figure}
The   stress-strain response,  following   the first instability at the point   A',   is illustrated in Fig. \ref{fig2_4_2}(a) for a particular choice of parameters $\lambda_1$ and $\lambda_2$. The 
branch switching  transition A'$\to$A  brings the system from the trivial branch with $n=0$ to  the nontrivial  equilibrium branch with $n=3$, and the linear stability of the ensuing  non-affine  configuration is illustrated in Fig. \ref{fig2_4_2}(b)  where we show that in  point A the smallest eigenvalue of the corresponding second variation is positive. In   Fig. \ref{fig2_4_2}(c,d) we show  that during this   symmetry breaking transition  two localized cracks  nucleate simultaneously: one inside the domain and one on the boundary.  Due to the  subcritical nature of the bifurcation,  the  dissipative transition  A'$\to$A  is accompanied by an abrupt  stress drop, see Fig. \ref{fig2_4_2}(a).  
\begin{figure}[h!]
\centering
\includegraphics[scale=.45]{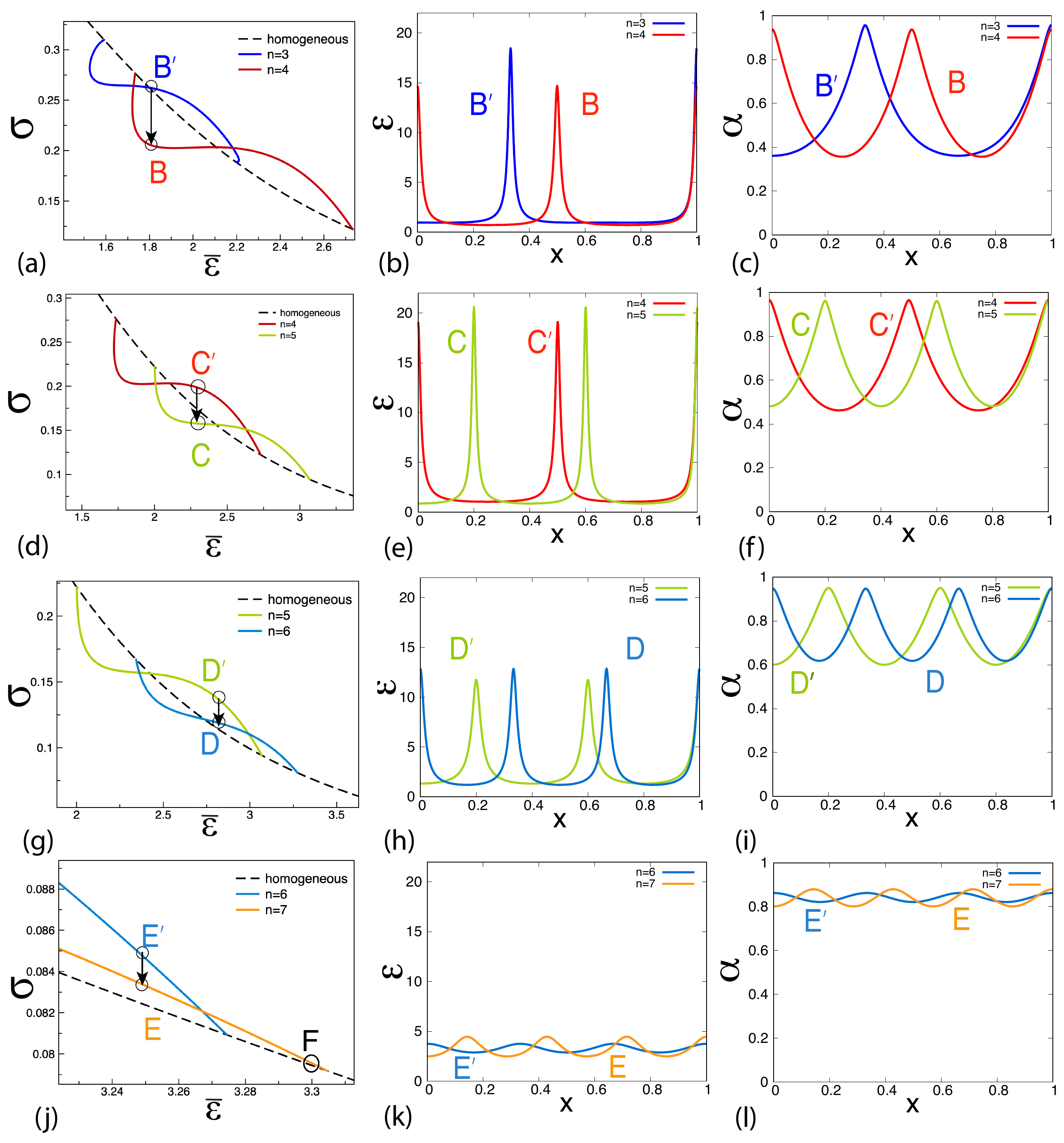}
\caption{Zoom in on various dissipation-free GEM transitions depicted in Fig. \ref{fig2_4}:   (a,d,g,j)    stress-strain relations,     (b,e,h,k)    strain profiles, (c,f,i,l)  damage parameter profiles.    Parameters: $\lambda_1=0.158114$, $\lambda_2=0.34$. }
\label{fig22}
\end{figure}
As we see from Fig.~\ref{fig2_4_2}(b), the equilibrium configurations with $n=3$ loses linear stability at the point A''. To describe the subsequent transformations, we need to reconstruct the global picture and consider equilibrium configurations with different values of $n$. The corresponding solutions of the nonlinear system \eqref{modeld_elf} are shown in Fig. \ref{fig2_4} where we collected information about  the   equilibrium branches   with $n=3,...,7$.

We observe   that under the LEM protocol, the only available transition from point A'' is to the branch with $n=4$, which is locally stable in the corresponding range of applied strains $\bar \eps$. Beyond this point, a sequence of dissipative LEM transitions takes place with more and more cracks appearing sequentially, till finally, at a sufficiently large value of the loading parameter, the strain localization abruptly disappears, and the damage becomes uniformly distributed. The detailed picture of the LEM response will be laid out elsewhere, while below, for a difference, we summarize the system's behavior along  the GEM dynamic path.

The access to the equilibrium branches shown in Fig.~\ref{fig2_4}  allow one to describe  all  the successive branch switching events  in the process of   quasi-static stretching. The nature of  the corresponding transitions   is illustrated in Fig.~\ref{fig22}  which should be compared with the analogous  Fig.~\ref{globalminmech} and Fig.~\ref{globalminmech1}  showing the  succession of the GEM transitions in the   system reinforced by pantograph substructure.

We observe in Fig.~\ref{fig2_4} that along the GEM path the first non-symmetric, two-crack configuration  nucleates non-dissipatively at the point G'. The corresponding transition G'$\to$G  marks  the switch form the branch $n=0$ to the branch  $n=3$,  see Fig. \ref{fig2_4}(a).  The next GEM  transition at point B' to the branch  $n=4$ takes place before the corresponding linear instability limit, marked by the point A'' is reached,  see  Fig.~\ref{fig2_4}(a). The  configuration which emerges as a result of the  non-dissipative  transition B'$\to$B is illustrated in Fig.~\ref{fig22} (a-c). As we see, the non-symmetric two-crack configuration transforms into the symmetric three-crack configuration with one localized crack in the center and two localized cracks around the boundaries. The  next GEM  transition C'$\to$C breaks the symmetry again, creating  a three-crack configuration with two localized cracks inside and one localized crack on the boundary. Then, the symmetry is   recovered   during the transition D'$\to$D when the   four-crack  configuration emerges with two localized cracks inside and two localized cracks around the boundaries. Finally, after yet another  symmetry breaking transition E'$\to$E the affine configuration  is retrieved at  the point F.

Note that with each successive GEM transition, both the strain $\eps(x)$ and the measure of damage $\a(x)$ become less localized. In particular, just before the affine state is recovered, that sample appears almost unstressed with slight modulation of strain but with the level of damage almost uniformly \emph{high}. In such  configuration, which absorbs all the work of the loading device,  the elastic strain is   systematically replaced by the inelastic strain. Instead, along  the corresponding LEM path, the energy is dissipated  instead of being accumulated.

Note also that in the phase-field model,  even in the presence of an elastic background, cracks remain \emph{localized} almost all the way till they disappear  in the state  entirely dominated by the elastic background.  Their number increases with stretch,  however,  in contrast to the model of  pantograph-reinforced chain, the extended \emph{domains} of distributed damage do not appear.  
%Failure takes the form of a pattern  of smeared out but   isolated displacement discontinuities.
 In this sense, the reinforcement through the  elastic foundation \emph{is not} equivalent to the reinforcement through the bending dominated sub-structure. Therefore, the floppy  substructure is indeed  the crucial element of the proposed metamaterial  design.

\section{Conclusions}
While some natural materials break with the formation of a single macro-crack, other natural materials  exhibit multiple, almost diffuse macro-cracking. The difference between the two classes of material behavior is reflected in the nomenclature of \emph{fracture} and \emph{damage} mechanics. The two  are often presented  as  separate disciplines addressing   fundamentally distinct failure modes; a closely related antithesis is between brittle and quasi-brittle (ductile) responses. 

Various attempts have been made to explain the difference between these two  failure mechanisms by linking them, for instance, to preexisting defects \cite{Shekhawat2013-qg,Borja_da_Rocha2020-vf}
or to the convexity properties of the cohesive energy \cite{Del_Piero2001-lb,Rinaldi2013-sk}.  Under the assumption that "to understand, is to build", we posed in this paper  the problem of designing an artificially engineered metamaterial that can be potentially switched from one of these failure modes to another.

Our main idea    is that the range of stress redistribution, exemplified  by the effective rigidity,  may serve as the  factor affecting, at least in some cases,  the localization properties of fracture phenomenon \cite{Driscoll10813}. In particular, we conjectured that the transition from `stretching dominated' to `bending dominated' elasticity \cite{Giessen:2011aa,buxton2007,Salman2019-uy} will favor strain delocalization and will be able to change the character of the cracking process from brittle-like to ductile-like.

 To check this possibility, we followed various earlier insights and  proposed  the simplest conceptual design of a high-toughness, pseudo-ductile metamaterial with nominally brittle sub-elements. Using  this toy example we showed that by affecting the nature of the structural connectivity inside an elastic system, one can transform a brittle structure, which fails with the formation of highly localized cracks, into an apparently ductile structure exhibiting de-localized damage. The desired nominal ductility is achieved by elastic coupling of a conventional stretching-dominated brittle sub-structure with another floppy sub-structure that can transmit bending-dominated nonlocal elastic interactions.
 
To substantiate these conclusions, we solved a series of elementary 1D model problems showing how the presence of a floppy sub-structure can suppress strain localization and induce the formation of diffuse zones of microcracking. To facilitate  the analysis we developed an asymptotically equivalent continuum theory of Ginzburg-Landau  type with strain as the order parameter. Since the local part of the corresponding   energy is represented by a \emph{single-well} potential with sub-linear growth, the nonlocal (gradient) term becomes relevant `volumetrically’ even though there is a small coefficient in front of it. This is unusual, given that in the conventional theory of phase transitions, a similar term is only relevant  for the description of narrow transition zones. We showed that in tensile loading, the proposed Ginzburg-Landau elastic model reproduces  the behavior of the original discrete model adequately, including the intriguing re-entrant \emph{isola-center } bifurcation. The main lesson is that the re-stabilization of the affine response can be  accomplished by bending rather than stretching elasticity.
\begin{figure}[h!]
\centering
\includegraphics[scale=.2]{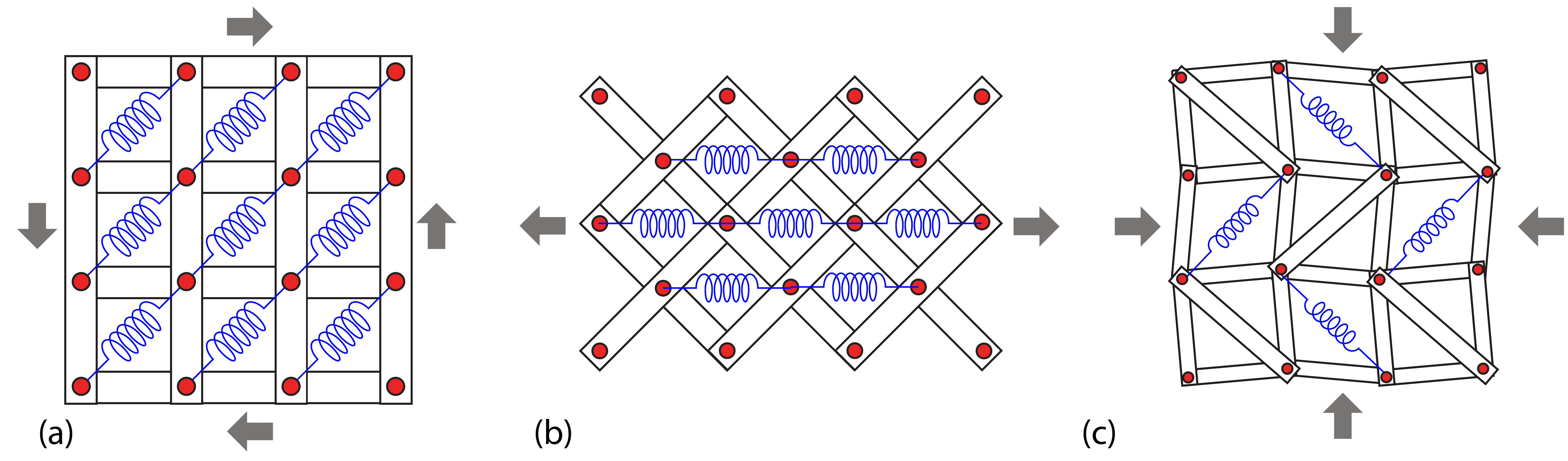}
\caption{Three examples of  the 2D   spring systems reinforced by floppy substructures with bending dominated elasticity that can be expected   to exhibit  fracture de-localization under: (a) shear, (b) uni-axial tension, (c) bi-axial (hydrostatic) compression.}
\label{fig:proto}
\end{figure}

The analysis of the same system coupled to an elastic background  revealed a complex succession of fracture patterns.  Depending on the presence or absence of the floppy reinforcing substructure, the emerging microstructures   include either diffuse zones of microcracking or isolated macrocracks. Our ability to manipulate such patterns in specially designed metamaterials can mimic living cells' ability to assemble and dis-assemble various load-carrying `frames' that are fine-tuned to match the  particular types of loading. The presence in the space of the loading parameters of a finite range where the mechanical response of the system is non-affine may be of interest to industrial applications. For instance, it implies that under monotone driving, the appropriately designed metamaterial can produce a transient, information-carrying failure pattern that first comes out but then gets erased. 

The proposed \emph{prototypical} design of the pantograph-reinforced mass-spring chain serves only as a proof of concept, and the technologically relevant 3D brittle metamaterials, reinforced by bending-dominated floppy networks, would still have to be designed and fabricated. This task, however,  is not unrealistic, given  the  already existing 3D printing capabilities which open access to   high-contrast composite networks with inextensible but bendable elements. Three potentially interesting 2D designs of this type, involving   \emph{floppy} substructures with bending dominated elasticity, and expected to show fracture delocalization  in either  shear,   uni-axial tension or bi-axial (hydrostatic) compression, are shown in Fig. \ref{fig:proto}. They demonstrate  how the  \emph{harnessed} floppiness  can be used to achieve high toughness in low-weight structures.
 
Future work in the proposed direction must also include the account of the irreversibility of damage which was downplayed in our analysis. Finite size effects were also largely neglected.  Interesting problems will be raised  by the development of rigorous finite strain continuum approximations in higher dimensions  accounting for both `local' and `nonlocal' sub-structures. The associated continuum problems are  of higher-order, requiring the development of new analytical and numerical approaches.

  \section{Acknowledgments}

The authors are grateful to G. Vitale  for  his various contributions to the initial version of this  paper. O. U. S. acknowledges helpful discussions with I. R. Ioanescu.  O. U. S. was supported  by  the grants ANR-19-CE08-0010-01, ANR-20-CE91-0010, and L. T. by the grant ANR-10-IDEX-0001-02 PSL.

%\bibliography{bending}

\end{document}